\documentclass[twocolumn]{aastex62}

\newcommand{\overden}{$\delta_{\rm LAE}$}
\newcommand{\optdep}{$\tau_{\rm LoS}$}
\newcommand{\lya}{Ly$\alpha$}

\usepackage{booktabs}
\usepackage{hyperref}
\usepackage{amsmath}
\usepackage{lineno}
\usepackage{graphicx}
\usepackage{graphbox}
\usepackage{enumitem}
\usepackage{bm}
\usepackage{color}
\usepackage[normalem]{ulem}

\graphicspath{{./}{figures/}}

\received{April 20, 2020}
\revised{\today}
\accepted{ZZZ}

\shorttitle{Stat Correlation between LAEs \& IGM {\sc Hi} at $z\sim2.2$}
\shortauthors{Liang et al.}

\begin{document}

\title{Statistical correlation between the distribution of Ly$\alpha$ emitters and IGM {\sc Hi} at z$\sim$2.2 \\ mapped by Subaru/Hyper Suprime-Cam}

\correspondingauthor{Yongming Liang}
\email{ym.liang@grad.nao.ac.jp}
\email{zcai@mail.tsinghua.edu.cn}

\author[0000-0002-2725-302X]{Yongming Liang}
\affil{Department of Astronomical Science, SOKENDAI (The Graduate University for Advanced Studies), Mitaka, Tokyo 181-8588, Japan}
\affil{National Astronomical Observatory of Japan, 2-21-1 Osawa, Mitaka, Tokyo 181-8588, Japan}
\affil{Department of Astronomy, School of Science, The University of Tokyo, 7-3-1 Hongo, Bunkyo-ku, Tokyo, 113-0033, Japan}

\author[0000-0003-3954-4219]{Nobunari Kashikawa}
\affil{Department of Astronomy, School of Science, The University of Tokyo, 7-3-1 Hongo, Bunkyo-ku, Tokyo, 113-0033, Japan}
\affil{National Astronomical Observatory of Japan, 2-21-1 Osawa, Mitaka, Tokyo 181-8588, Japan}

\author[0000-0001-8467-6478]{Zheng Cai}
\affil{Department of Astronomy, Tsinghua University, Beijing 100084, China}
\affil{UCO/Lick Observatory, University of California, 1156 High Street, Santa Cruz, CA 95064, USA}

\author[0000-0003-3310-0131]{Xiaohui Fan}
\affil{Steward Observatory, University of Arizona, 933 N. Cherry Avenue, Tucson, AZ 85721, USA}

\author[0000-0002-7738-6875]{J. Xavier Prochaska}
\affil{UCO/Lick Observatory, University of California, 1156 High Street, Santa Cruz, CA 95064, USA}

\author[0000-0002-2597-2231]{Kazuhiro Shimasaku}
\affil{Department of Astronomy, School of Science, The University of Tokyo, 7-3-1 Hongo, Bunkyo-ku, Tokyo, 113-0033, Japan}

\author{Masayuki Tanaka}
\affil{Department of Astronomical Science, SOKENDAI (The Graduate University for Advanced Studies), Mitaka, Tokyo 181-8588, Japan}
\affil{National Astronomical Observatory of Japan, 2-21-1 Osawa, Mitaka, Tokyo 181-8588, Japan}

\author{Hisakazu Uchiyama}
\affil{National Astronomical Observatory of Japan, 2-21-1 Osawa, Mitaka, Tokyo 181-8588, Japan}

\author[0000-0002-9453-0381]{Kei Ito}
\affil{Department of Astronomical Science, SOKENDAI (The Graduate University for Advanced Studies), Mitaka, Tokyo 181-8588, Japan}
\affil{National Astronomical Observatory of Japan, 2-21-1 Osawa, Mitaka, Tokyo 181-8588, Japan}
\affil{Department of Astronomy, School of Science, The University of Tokyo, 7-3-1 Hongo, Bunkyo-ku, Tokyo, 113-0033, Japan}


\author[0000-0003-4442-2750]{Rhythm Shimakawa}
\affil{National Astronomical Observatory of Japan, 2-21-1 Osawa, Mitaka, Tokyo 181-8588, Japan}

\author[0000-0001-7457-8487]{Kentaro Nagamine}
\affil{Department of Earth and Space Science, Osaka University, 1-1 Machikaneyama, Toyonaka, Osaka 560-0043, Japan}
\affil{Kavli IPMU (WPI), The University of Tokyo, 5-1-5 Kashiwanoha, Kashiwa, Chiba 277-8583, Japan}
\affil{Department of Physics and Astronomy, University of Nevada, Las Vegas, 4505 S. Maryland Pkwy, Las Vegas, NV 89154-4002, USA}

\author{Ikkoh Shimizu}
\affil{National Astronomical Observatory of Japan, 2-21-1 Osawa, Mitaka, Tokyo 181-8588, Japan}
\affil{Shikoku Gakuin University, 3-2-1 Bunkyocho, Zentsuji, Kagawa, 765-8505, Japan}

\author{Masafusa Onoue}
\affil{Max-Planck-Institut f\"{u}r Astronomie, K\"{o}nigstuhl 17, D-69117 Heidelberg, Germany}

\author[0000-0001-5394-242X]{Jun Toshikawa}
\affil{Institute for Cosmic Ray Research, The University of Tokyo, Kashiwa, Chiba 277-8582, Japan}
\affil{Department of Physics, University of Bath, Claverton Down, Bath, BA2 7AY, UK}



\begin{abstract}

The correlation between neutral Hydrogen ({\sc Hi}) in the intergalactic medium (IGM) and galaxies now attracts great interests.
We select four fields which include several coherently strong \lya~absorption systems at $z\sim2.2$ detected by using background quasars from the whole SDSS/(e)BOSS database.
Deep narrow-band and $g$-band imaging are performed using the Hyper Suprime-Cam on the Subaru Telescope.
We select out 2,642 \lya~emitter (LAE) candidates at $z=2.177\pm0.023$ down to the \lya~luminosity of $L_{\text{Ly}\alpha}\approx 2 \times 10^{42}~{\rm erg~s}^{-1}$ to construct the galaxy overdensity maps, covering an effective area of 5.39 deg$^2$. 
Combining the sample with the \lya~absorption estimated from 64 (e)BOSS quasar spectra, we find a moderate to strong correlation between the LAE overdensity \overden~and the effective optical depth \optdep~in line-of-sights, with $P$-value$=0.09\%$ ($<0.01\%$) when the field that contains a significant quasar overdensity is in(ex)cluded.
The cross-correlation analysis also clearly suggests that 
up to $4\pm1$ pMpc, 
LAEs tend to cluster in the regions rich in {\sc Hi} gas, indicated by the high \optdep, and avoid the low \optdep~region where the {\sc Hi} gas is deficient.
By averaging the \optdep~as a function of the projected distance ($d$) to LAEs, we find a $30\%$ excess signal at $2\sigma$ level at $d<200$ pkpc, indicating the dense {\sc Hi} in circumgalactic medium, and a tentative excess at $400<d<600$ pkpc in IGM regime, corroborating the cross-correlation signal detected at about $0.5$ pMpc.
These statistical analyses indicate that galaxy$-$IGM {\sc Hi} correlations exist on scales ranging from several hundred pkpc to several pMpc at $z\sim2.2$.

\end{abstract}



\section{Introduction} 
\label{sec:intro}

The gravitational instability leads mass to assemble in a hierarchical manner from a uniform phase in the early universe, and galaxy formation occurs preferentially along large-scale filamentary and sheet-like overdense regions where the neutral Hydrogen ({\sc Hi}) in the intergalactic medium (IGM) is more abundant \citep{Springel+2006, Baugh+2006, Hinshaw+2007}.
The intersections of such filaments or sheets then evolve into dense clusters of galaxies at a later epoch \citep{Bond+1996, Cen+2000}. 
Therefore, the overdensities at high-$z$ are the crucial laboratories to study the large-scale structure (LSS) formation and evolution, especially 
the correlation between galaxy and IGM {\sc Hi}.

However, it is not easy to find the overdense regions at $z>2$, which only occupies a small fraction of the cosmic volume, e.g., $<2\%$ for protoclusters \citep{Chiang+2017}.
To make efficient surveys for galaxies, some studies use galaxies with radio loud active galactic nucleus (AGNs) \citep{Cooke+2014, Shimakawa+2014, Noirot+2018}, dusty star forming galaxies \citep{Casey+2015}, luminous quasars \citep{Kikuta+2019} or quasar pairs \citep{Onoue+2018} as overdensity tracers.
Because such rare objects are expected to reside in massive halos, which are likely to host the protoclusters.  
In addition, damped \lya~systems (DLAs) \citep{Ogura+2017, Fumagalli+2017} or systems with extended nebular emission around galaxies \citep{Badescu+2017} are the good candidates as tracers as well.
Wide-field surveys also enable blind searches of protoclusters via photo-z galaxies \citep{Spitler+2012} and Lyman-break galaxies (LBGs) \citep{Toshikawa+2016, Toshikawa+2018}. 

In addition to galaxy surveys, for decades in simulations, the LSSs in terms of IGM {\sc Hi} have also been demonstrated to be possibly revealed by the absorptions imprinting in the spectra of background quasars \citep{Hernquist+1996, Springel+2006}, and it is also proved to be a non-trivial question at high-$z$ universe, as most baryons at $z>2$ may reside in \lya~clouds \citep{ME+1996}.
Strong {\sc Hi} absorbers are studied around quasars \citep{Prochaska+2013} or with searching the associated galaxies \citep{Mackenzie+2019}, from which a hint of the galaxy-IGM {\sc Hi} correlation is found.
Based on a specific field SSA22 with the protocluster found at $z=3.1$, \citet{Mawatari+2017} have found a global correlation on a scale of tens of comoving-Mpc (cMpc) via the narrowband absorption technique. 
\citet{Hayashino+2019}, who study the same structure, find the similar correlation in the redshift space.

The galaxy-IGM {\sc Hi} correlation can also be studied in a statistical way with large galaxy surveys for foreground LBG/photo-$z$ galaxy and the background quasar/galaxy pairs \citep{Adelberger+2003, Adelberger+2005, Steidel+2010, Rudie+2012, Turner+2014, Mukae+2017, Momose+2020b, Chen+2020}, and most of these researches find the correlation on various scales. 
However, these studies are confined by either the bright galaxy populations, or the relatively small dynamic range of the IGM absorption due to the limited sample size and survey area.

Recently, the IGM tomography also becomes feasible to construct 3D IGM {\sc Hi} maps from the background star-forming galaxies \citep{Lee+2014a, Lee+2014b, Lee+2016, Lee+2018, Newman+2020}.
But, the tomography surveys to date are still limited by the survey area $\lesssim1$ deg$^2$, and it is mainly conducted on blank fields.
A larger survey area covering various overdense regions is essential to take full advantage of the technique.

In the MApping the Most Massive Overdensity Through Hydrogen (MAMMOTH) project \citep{Cai+2016, Cai+2017a, Cai+2017b}, N-body simulations suggest that coherently strong \lya~absorption system (CoSLAs), originated from the overlapping of the Ly$\alpha$ forest, can effectively trace the most massive halos on the scale over 15 $h^{-1}$cMpc.
Although whether CoSLAs traces well the most massive overdensity is now under debate \citep{Miller+2019}, a pilot MAMMOTH program has found the BOSS1441, one of the most massive structures to date at $z>2$, with also six BOSS quasars associated \citep{Cai+2017a}. 

While helping to pinpoint the regions that tend to host overdensities, the grouping rare lines-of-sight (LoSs) with high IGM absorption (e.g., CoSLAs) also enable us to significantly enhance the dynamic range in statistics for studying the galaxy-IGM {\sc Hi} correlation.
Targeting the fields centered at MAMMOTH candidates on 15 $h^{-1}$cMpc, the Subaru/Hyper Suprim-Cam  \citep[HSC;][]{Miyazaki+2018} equipped with a diameter $d=1.5$ deg field-of-view (FoV) makes it possible to efficiently map the most diverse universe at $z\sim2$ on a scale over 100 cMpc.
Additionally, the narrowband technique for identifying \lya~emitters (LAEs) whose redshifts can be well constrained in a narrow range ($\Delta z<0.05$), also opens a window towards a fainter and less massive galaxy population for the correlation.



In this paper, we first summarize the SDSS/(e)BOSS data, the field selection, the Subaru/HSC observations and the data processing in Section \ref{sec:data}. 
The LAE sample construction are then presented in Section \ref{sec:sample}. 
Section \ref{sec:result} shows our LAE overdensity maps for the four HSC fields. 
The analyses of the galaxy-IGM {\sc Hi} correlation are also shown in this section.
Section \ref{sec:discuss} compares our results with other works, and explores the scale dependence of the correlation.
The underlying physics is also discussed in the last part.
Finally, we end with a summary and give an outlook of the future work in Section \ref{sec:summary}.
The cosmological parameters used in this paper are based on \citet{Planck15XIII}: $H_0=67.7~{\rm km~Mpc}^{-1}{s}^{-1}$, $\Omega_0 = 0.307$.
{\it AB} magnitudes are used throughout the paper.

\section{Data}
\label{sec:data}

\subsection{SDSS/BOSS Spectral Data}
\label{sec:boss_spec}
The background quasar spectra from the Baryon Oscillation Spectroscopic Survey (BOSS) of SDSS-III \citep{Dawson+2013} and the later upgraded extended-BOSS, or eBOSS, of SDSS-IV \citep{Dawson+2016} are used in this work for both selecting candidate fields and evaluating the effective optical depth in the correlation analysis.
BOSS is a spectroscopic survey specially designed to study the intergalactic science through Ly$\alpha$ forest.
It takes spectra with the 2.5-m Sloan telescope for over 150,000 background quasar at $z\gtrsim2.15$ reaching a depth as faint as $g<22$. 
The eBOSS observes 60,000 BOSS quasars for spectra in better quality and 60,000 new targets in complement. 
The surveys combined offer more than 200,000 quasar spectra covering a survey area of over 10,000 deg$^2$, corresponding to a survey volume of $>$ 1 Gpc$^3$.

The (e)BOSS database offers us abundant quasar spectra working as LoSs, in which the IGM distribution can be traced by the \lya~absorption.
To evaluate the \lya~absorption, we calculate the effective optical depth in the LoS, \optdep, within the \lya~redshift range traced by the narrowband filter NB387 ($\lambda_0$ = $3,862$ \AA, FWHM = 56 \AA).

We first smooth the flux along the wavelength dimension over a scale of 15 $h^{-1}$ cMpc.
Absorption features are searched by scanning through the spectra over a range of $\pm35$ \AA~centered around $3,862$ \AA.
The effective optical depth is then calculated at the strongest absorption spike following \citet{Cai+2016}:
\begin{equation}
    \tau_{\rm LoS} = - \ln \left< F\right>_{15\,h^{-1}{\rm cMpc}},
    \label{eq:tau_def}
\end{equation}
where the $\left< F \right>_{15\,h^{-1} {\rm cMpc}}$ is the continuum normalized flux estimated on the $15\,h^{-1} {\rm cMpc}$ scale.
Note that the \optdep~estimated here can be systematically larger than the cosmic mean, as we are targeting at IGM \lya~absorbers as the gas tracers, instead of the random forest.

When evaluating the \lya~absorption by using the (e)BOSS spectra, quasar continuum is constructed using the mean-flux-regulated principal component analysis (MF-PCA) technique to the fitting \citep{Lee+2013}. 
The extra constraints on the slope and amplitude of the continuum are adjusted by using the mean optical depth of the \lya~forest \citep{Lee+2012, Becker+2013}.

The \optdep~will be used throughout the paper for both the field selection and the galaxy-\lya~absorption correlation analysis.

\subsection{Field Selection}
\label{sec:field_select}

Our goal in this paper is to study the galaxy-IGM {\sc Hi} correlation on a wide range of the environments based on the less massive galaxy populations.
The principle for our field selection is 
to enclose a sufficient number of LoSs, especially those with strong \lya~absorptions, while we also target the possible overdensities.

First, we briefly summarize the selection of COSLAs, but please see the details in \citet{Cai+2016}.
The LoSs with $\tau_{\rm LoS} \gtrsim 3\left< \tau\right>_{cos}$ is chosen as the preliminary absorber candidates, where the $\left< \tau\right>_{cos}$ is the cosmic mean optical depth, and we assume it as 0.15 at $z=2.2$ \citep{Becker+2013} with slight adjustments according to fields.
To eliminate the non-IGM contaminants, we make the systematic inspections of the criteria proposed in \citet{Cai+2016} to reject the high column density systems (HCDs), i.e., DLAs, sub-DLAs or Lyman-limit systems (LLS), and we also do the visual checks to remove the broad absorption line (BAL) quasars that may confuse the interpretation of IGM \lya~absorption in the NB387 wavelength range.
Besides for the high \optdep~LoSs, all of these processes for excluding the non-IGM contaminants are also done for the potential LoSs used in our following analysis, which we call as the clean LoSs here.

Therefore, based on the clean LoSs, we own several preferences when selecting the target fields:
the target fields of HSC-FoV should:
(1) contain the high \optdep~LoSs to expand the dynamic range;  
(2) enclose as many LoSs as possible to increase sample size for drawing the galaxy-IGM {\sc Hi} relation;
(3) contain a concentration of high \optdep~LoSs to find a protocluster, i.e., $\gtrsim 4$ LoSs within a ($\sim20~h^{-1}$cMpc)$^3$ box, which is the typical scale of a Coma-type protocluster at the $z\sim2$ \citep{Chiang+2013};
(4) in special case, contain the associated quasars at $2.15 \le z \le 2.20$, i.e., proximity quasars at the similar redshift of our LAEs, to see any 
possible difference.

Our field selection were, however, further compromised by the field visibility, the distance to the moon or the nearby bright stars in a specific observation run.
As a result, four fields BOSS$J0210+0052$ (or $J0210$), BOSS$J0222-0224$ (or $J0222$), BOSS$J0924+1503$ (or $J0924$) and BOSS$J1419+0500$ (or $J1419$) are selected in our observations, and all of them satisfy (1) and mostly (2). 
$J0222$ and $J0924$ are selected mainly based on (3), the typical regions hinting the coherent IGM {\sc Hi} on large scale.
$J1419$ was once selected also for (3), but one of the two concentrating LoSs, which is found to be a possible BAL quasar\footnote{
This is J141934.64+050327.1, which is categorized as a {\it probable} {\sc Pv} BAL quasar in \citet{Capellupo+2017}} after the observation, is excluded from our analysis.
However, the field is still one of the best candidates considering (1) and (2), although the coherent IGM \lya~absorption is not as significant as other selected fields here.
Specially, $J0210$ is selected with the consideration of (4), given that a group of 11 proximity quasars is associated within a region of 40 $\times$ 40 cMpc$^2$ at $2.15<z<2.20$, a length of $62$ cMpc along the LoS direction, which is more extreme than the BOSS1441 found in \citet{Cai+2017a}.
One of the proximity quasars also shows the hint of strong IGM \lya~absorption at the wing of \lya~emission, but being conservative, we do not include it in our correlation analysis.
The coordinates of the field centers are listed in the Table \ref{tab:field}.

\begin{figure}[hbt!]
    \centering
    \includegraphics[width=\linewidth]{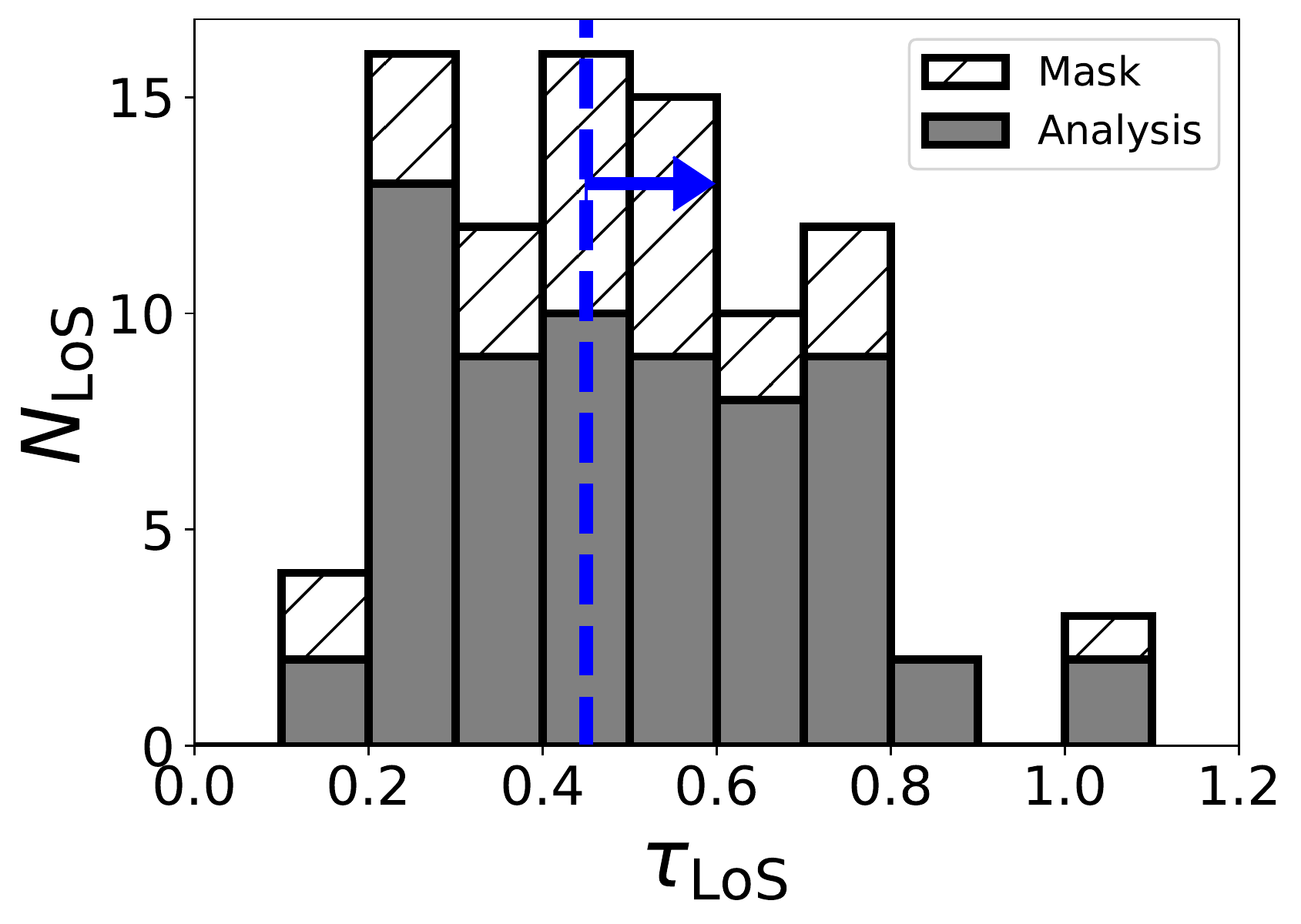}
    \caption{The \optdep~distribution of all inspected clean LoSs. 
    The hatched area indicates the masked LoSs, while the filled area shows the LoSs
    used in the following correlation analysis.
    The blue vertical dash line suggests the \optdep~criterion of the clean LoSs with \optdep~$\gtrsim 3 \left< \tau \right>_{cos}$.}
    \label{fig:hist_tau}
\end{figure}

We note that before applying mask in the following sections, there are 26, 23, 19 and 22 clean LoSs in $J0210$, $J0222$, $J0924$ and $J1419$ respectively, and they are summarized in Table \ref{tab:lae}.
The \optdep~distribution of these clean LoSs is shown in Figure \ref{fig:hist_tau}, in which a vertical blue dash line indicates the criterion for the clean LoSs with \optdep~$\gtrsim 3 \left< \tau \right>_{cos}$.

\begin{table*}[t!]
\scriptsize
\hspace*{-2.2cm}
\begin{tabular}{@{}ccccccccc@{}}
\toprule
Field & RA (J2000) & DEC (J2000) & Obs Period & FWHM$_{\rm PSF, NB}$ & FWMH$_{{\rm PSF}, g}$ & $m_{\rm NB, 5\sigma}$ & $m_{g, 5\sigma}$ & {\it hscPipe}  \\    \midrule
BOSS J0210+0052 & 02:09:58.90 & +00:53:43.0 & Jan., 2018       & 1.22\arcsec            & 0.90\arcsec           & 24.36\tablenotemark{a}                  & 26.24\tablenotemark{a}            & v5.4        \\
                &             &             &                 & 1.22\arcsec            & 0.90\arcsec           & 24.25\tablenotemark{a}                  & 26.34\tablenotemark{a}            & v6.6        \\
BOSS J0222-0224 & 02:22:24.66 & -02:23:41.2 & Jan., 2018       & 0.90\arcsec            & 0.90\arcsec           & 24.99                                   & 27.01                             & v5.4             \\
BOSS J0924+1503 & 09:24:00.70 & +15:04:16.7 & Jan. \& Mar., 2019  & 0.84\arcsec           & 0.79\arcsec           & 24.74                  & 26.63                & v6.6             \\
BOSS J1419+0500 & 14:19:33.80 & +05:00:17.2 & Mar., 2019       & 0.86\arcsec            & 0.70\arcsec           & 24.81                 & 26.80              & v6.6            \\ \bottomrule
\end{tabular}
\tablenotetext{a}{Measured in 2.5$^{\prime\prime}$ aperture, and $g$-band is the PSF-matched image.}
\caption{Summary of field information. 
Column 1 is the full name of fields;
Columns 2 and 3 are the coordinates RA and DEC in equinox with an epoch of J2000;
Column 4 is the period of executing the observations;
Columns 5 and 6 are the FWHMs of star PSFs measured for the final stacked images of NB387 and $g$-band;
Columns 7 and 8 are the 5$\sigma$ limiting magnitudes measured in an aperture with the radius of 1.7\arcsec for the final stacked NB387 image and PSF-matched $g$-band, respectively; 
Column 9 is the {\it hscPipe} version used for the data reduction.}
\label{tab:field}
\end{table*}

\begin{table*}[t!]
\centering
\begin{tabular}{@{}ccccccc@{}}
\toprule
Field & N$_{\rm LoS, All}$ & N$_{\rm LoS, Ana}$ & N$_{\rm LAE}$ & Area [deg$^2$] & E(B-V)  \\ \midrule
BOSS J0210+0052 & 26 & 22 & 465 & 1.34  & 0.0246 \\
BOSS J0222-0224 & 23 & 11 & 956 & 1.13  & 0.0222           \\
BOSS J0924+1503 & 19 & 14 & 585 & 1.47  & 0.0217        \\ 
BOSS J1419+0500 & 22 & 17 & 636 & 1.45  & 0.0264       \\
\midrule
Total & 90 & 64 & 2642 & 5.39 & \textbackslash{}  \\
\bottomrule
\end{tabular}
\caption{Information of LoSs and LAEs in each field. 
Column 1 is the respective field; 
Column 2 is the number of all the clean LoSs inspected in/around the four fields;
Column 3 is the number of the clean LoSs after being masked, which are used in the correlation analysis in this work; 
Column 4 is the number of LAE candidates; 
Column 5 is the effective survey area for selecting LAEs after being masked; 
Column 6 is the galactic reddening accounting the Milky Way based on the measurement and calibration from \citet{SF11}.}
\label{tab:lae}
\end{table*}

\subsection{Imaging Observations}
\label{sec:obs}

Observations to identify LAEs were carried out with the HSC installed at the prime focus of the 8.2-m Subaru telescope located at the summit of the Mauna Kea, Hawaii. 
HSC is a high performance camera with a wide FoV of $1.5$ deg in diameter. 
As a gigantic mosaic CCD camera, HSC consists of 104 Hamamatsu Photonics KK CCDs (2048 $\times$ 4096 pixels) for science, 4 for auto-guiders and 8 for focusing. 
The pixel scale of the CCD reaches 0.168\arcsec. 

\begin{figure}[hbt!]
    \centering
    \includegraphics[width=3.2in]{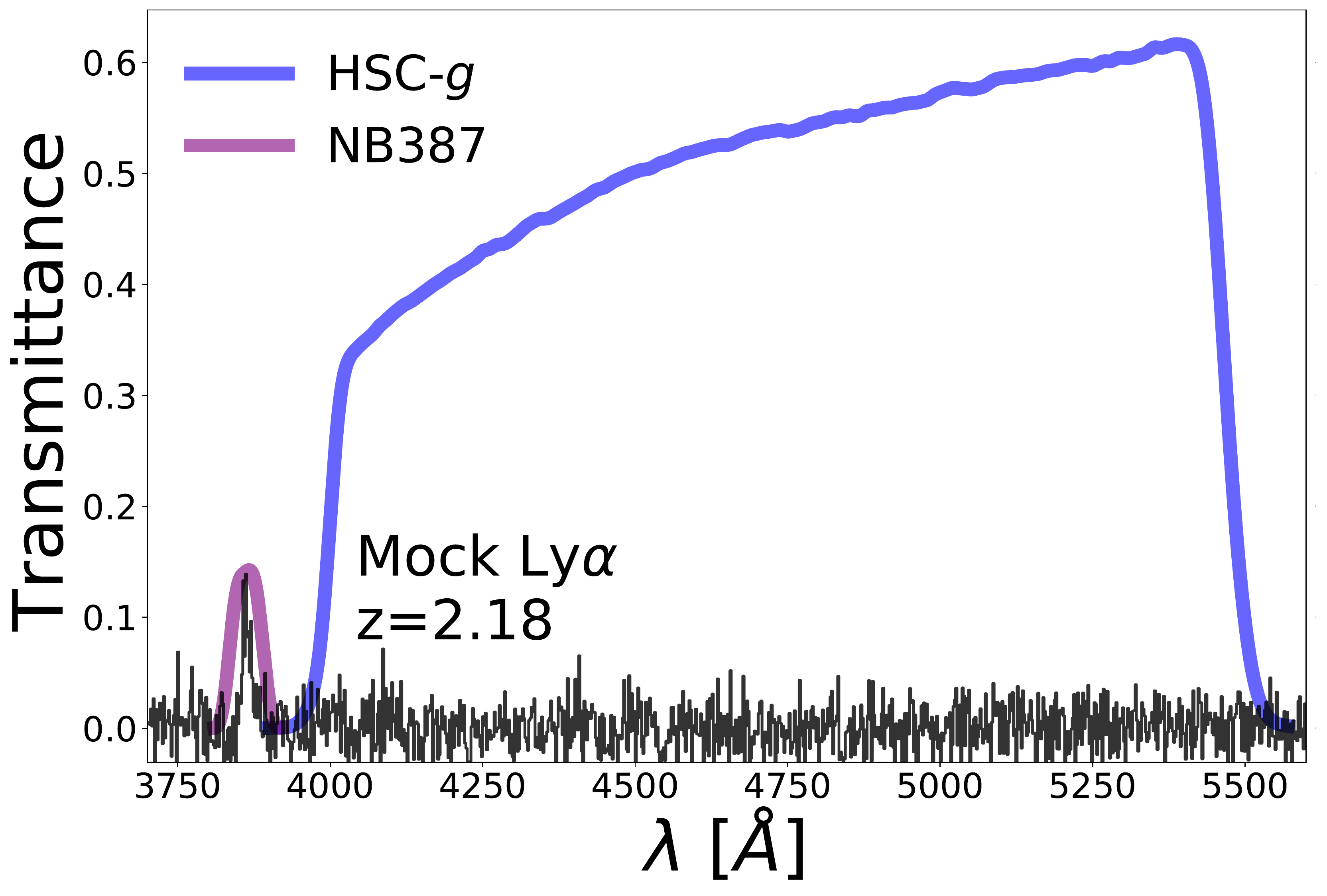}
    \caption{Transmission curve of HSC-$g$ and NB387 band. 
    The purple and blue solid lines are the total transmittance of NB387 and HSC-$g$ counting the CCD quantum efficiency, dewar window, the Primary Focus Unit and the reflectivity of the Prime Mirror. 
    The black curve indicates a mock LAE spectrum at $z=2.18$, whose \lya~emission is exactly located at the sensitive wavelength range of the NB387.}
    \label{fig:filter_trans}
\end{figure}

In this paper, we perform the deep NB imaging using the NB387, which enables us to detect \lya~emission at the corresponding redshift of $z=2.177 \pm 0.023$. 
The $g$-band is also used for the evaluation of the continuum level of the detected objects. 
The transmission curves of the filters, which has taken the transmittance accounting in CCD quantum efficiency, dewar window, the Primary Focus Unit and the reflectivity of the Prime Mirror into account, are shown in Figure \ref{fig:filter_trans}. 
To ensure the depth for detecting a sufficient number of LAEs, the observation is designed to have total exposures of 3 hours for NB387 and 40 min for $g$-band in each field. 
An S17B observation was carried out in a queue mode in Jan. 2018 and Jan. 2019, and exposures are split into 900 s and 600 s for the NB387 and $g$-band respectively, except for the first 2 exposures of 1,200~s for $J0210$. 
In the S19A observation carried out on-site on Mar. 8$^{\rm th}$, 2019, the exposures are split into 900s and 300s for NB387 and $g$-band respectively to avoid saturations in the broadband. 
From S17B, we have obtained both the NB387 and $g$-band data for fields $J0210$ and $J0222$ and the $g$-band data for $J0924$. 
In the S19A run, NB387 data for both $J0924$ and $J1419$ are achieved, and the $g$-band observation is taken for $J1419$ only. 
In summary, both NB387 and $g$-band imaging data for all four fields are from two major runs. 

Except for the NB imaging of $J0210$ in relatively poor conditions with seeing over 1.2\arcsec, all of the observations were executed under moderate to good conditions. 
Some exposures are discarded because of the occasionally poor seeing or low transparency. 
In $J0222$, the severe stray light from a nearby Mira contaminates some exposures and they are also discarded. 
Standard stars are not used considering the large FoV and 104 CCDs.
Instead, we use Pan-STARS DR1 (PS1) photometric data \citep{Chambers+2016} for calibration as described in the Sec. \ref{sec:reduction}.
Detailed information of each field is summarized in the Table \ref{tab:field}.

\subsection{Data Reduction}
\label{sec:reduction}

The NB387 and $g$-band imaging data are reduced with the HSC pipeline, {\it hscPipe} \citep{Bosch+2017, Aihara+2019}. 
$J0210$ and $J0222$ are reduced with the {\it hscPipe} 5.4, and $J0924$ and $J1419$ are reduced with the {\it hscPipe} 6.6. 
Given the relatively poor quality of the $J0210$ NB387 data, we also reduce both the $g$-band and NB387 data for $J0210$ with $hscPipe$ 6.6 and combine the catalog with the one produced from $hscPipe$ 5.4. The overlaid detections with separations smaller than 2\arcsec~are only kept for the latter version.
\cite{Bosch+2017} and \cite{Aihara+2019} describe the data reduction process as well as the code updates in details, but we give a brief summary here with an emphasis on the processes that are different from the standard usage. 

The {\it hscPipe} first makes calibration data, including the bias, dark, dome-flat and global sky.
Then it applies them to each CCD in single visit, and a local sky background on 128 pixel scale is subtracted.
Bright objects are then extracted for astrometric and photometric calibration.
Point-spread function (PSF) models used inside the pipeline are also made at this step.
Astrometry and photometry are then calibrated against PS1 references.
For each filter, the zero-point is adjusted by fitting a multi-band relation, e.g., a ${\rm NB} - g~{\rm vs.}~g - r$ relation for the NB387:
\begin{equation}
    \begin{split}
    {\rm NB387}_{\rm HSC} - g_{\rm PS1} = &~0.541 \times \left( g_{\rm PS1} - r_{\rm PS1} \right)^2 \\
     & + 1.87 \times \left( g_{\rm PS1} - r_{\rm PS1} \right) \\
    & + 0.428~[+ C_{\rm metal} + C_{\rm fit}],
    \end{split}
    \label{eq:photo_check}
\end{equation}
which is derived from the template magnitudes predicted by spectroscopic Pickles star references \citep{Pickles+1998} and the filter transmissions.
A tract ID is defined to enclose all observed sky, and then a sky map is made as the reference for the following coadding process.
A global sky background was subtracted without masked regions.
In the mosaicking of the CCDs data, both the WCS and the flux scale were corrected by a spatially-varying correction term. 
Finally, the coadding process warp the images to the sky map and co-add all visits of the image together scaled with the WCS and flux correction from the mosaicking process. 

For our data, some configurations need to be further optimized in addition to the aforementioned process.
The NB387 image is always too shallow in a single frame image to have enough bright stars in each CCD for calibration.
Therefore, we set the parameter set for choosing calibration stars lower by $\sim0.5~\times$ default value. 
In addition, when fitting the Equation \ref{eq:photo_check} to determine the photometric zero-point of NB387, we take into accounts additional corrections, including a systematic correction $C_{\rm metal}$ of $0.448$ mag for correcting the stellar metalicity bias, and a field dependent term $C_{\rm fit}$ ranging within $0.2$ mag for calibrating the fitting uncertainties. 
Details are described in Appendix \ref{appendix:zp_correct}.

\subsection{Photometric Processing}
\label{sec:photometry}

We use the {\it SExtractor 2.19.5} \citep{Bertin+1996} for the photometry processing.
First, we do the PSF matching for the $g-$band and NB387 images by convolving a proper Gaussian kernel in each field. 
Then we run the dual-image mode for the source detection and measurement by setting the NB387 image as the reference. 
The detection threshold is set as 15 continuous pixels over the $1.2\sigma$ sky background. 
Because of the large HSC FoV and the mosaic CCD structure, there are slight fluctuations of 0.1 - 0.2 mag of the image depth among the whole field. 
We apply the sky background root-mean-square (RMS) map as the weighting map in {\it SExtractor} to minimize this influence. 
In addition, we use a local background with the thickness of 128 pixels.
Masks are also applied when doing the background estimation, object detection and measurement.
The masks are defined as regions with low S/N signals, saturation around bright stars or severe stray lights.

Note that after applying the masks, the final numbers of clean LoSs in $J0210$, $J0222$, $J0924$ and $J1419$ are 22, 11, 14 and 17 respectively, as summarized in the Table \ref{tab:lae}. 
The masked clean LoSs are hatched in Figure \ref{fig:hist_tau}, and the remaining 64 clean LoSs will be used for all of the following correlation analysis, unless some of them are further removed with the nearby masked regions over a certain fraction, as described in Section \ref{sec:od2_direct} and \ref{sec:od2_scale}.

We use aperture magnitudes for the color selection, and the aperture diameters are 15 pixels ($\sim2.5$\arcsec) for $J0210$ and 10 pixels ($\sim1.7$\arcsec) for $J0222$, $J0924$ and $J1419$.
The Auto-Mag is used for the estimate of total magnitude, which applies automatically determined elliptical aperture for Kron photometry in {\it SExtractor}.
Galactic extinction is also accounted in each band.
Referring to the Galactic Dust Reddening and Extinction Service provided by IRSA, which is based on the results of \cite{SF11}, color reddening $E(B-V)$ can be estimated and is listed in Table \ref{tab:lae}. 
As the $R_{\rm NB387} = A_{\rm NB387}/E(B-V)$ is estimated as 4.009, taking into account of the transmission curve, we manage to apply the dust extinction correction for the detection catalogs. 
We replace the $g$-band magnitude with the corresponding 2$\sigma$ limiting magnitude, when the objects are fainter than the 2$\sigma$ limit.

The measured PSF FWHM and the 5$\sigma$ limiting magnitudes in the 1.7\arcsec~aperture (2.5\arcsec~for $J0210$) of the final stacked images of NB387 and $g$-band are listed in the Table \ref{tab:field}. 
The quality of $J0210$ data is relatively poorer compared to the other three fields, in both of the seeing and the final image depth.

\section{Sample Selection} 
\label{sec:sample}

\begin{figure*}[hbt!]
    \centering
    \includegraphics[width=0.8\textwidth]{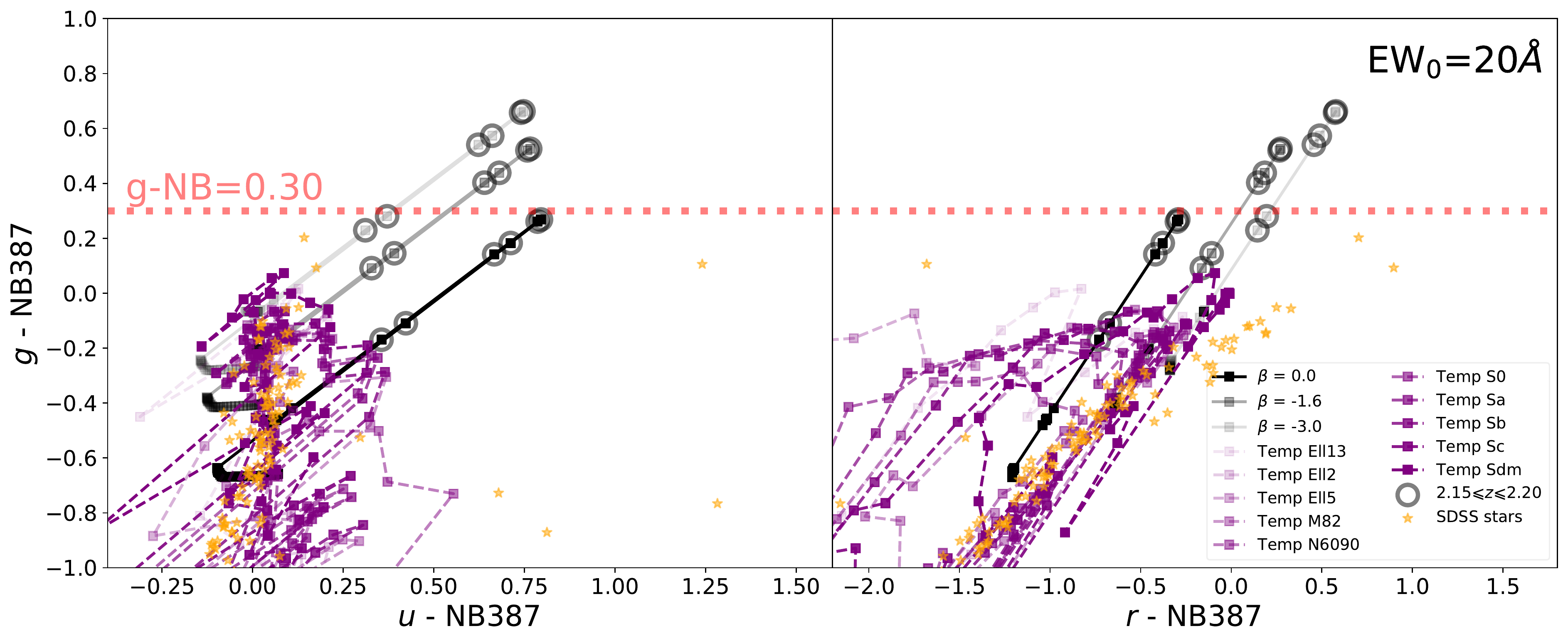}
    \caption{The color tracks with redshift evolution for the $EW_0 = 20$ \AA~LAE at the $z=2 - 2.5$: $g - {\rm NB387}$ vs. $u - {\rm NB387}$ ({\it left panel}) and $g - {\rm NB387}$ vs. $u - {\rm NB387}$ ({\it right panel}).
    The black curves are the tracks for LAE models (with UV slope $\beta=0, -1.6, -3.0$) with a redshift step $\Delta z=0.01$ from $z=2$ to $z=2.5$, and the purple dash curves are for galaxy templates, including elliptical galaxies (age of 2, 5 and 13 Gyr denoted as Ell2, Ell5 and Ell13), starburst galaxies (M82 and N6090) and spiral galaxies (S0, Sa, Sb, Sc, Sd and Sdm) \citep{Polletta+2007} with a redshift step $\Delta z=0.1$ from $z=0$ to $z=3$.
    The homogeneously picked SDSS stars with $g>19$ \citep{Yanny+2009} are also plotted as yellow stars.
    Circles indicate the LAE models at $2.15 \leqslant z \leqslant 2.20$.
    The narrowband excess $g-{\rm NB387} > 0.30$ works as a reasonable threshold to select out the $z \sim 2.18$ LAEs.}
    \label{fig:lae_select_criteria}
\end{figure*}

\begin{figure}[htb]
    \centering
    \includegraphics[width=\linewidth]{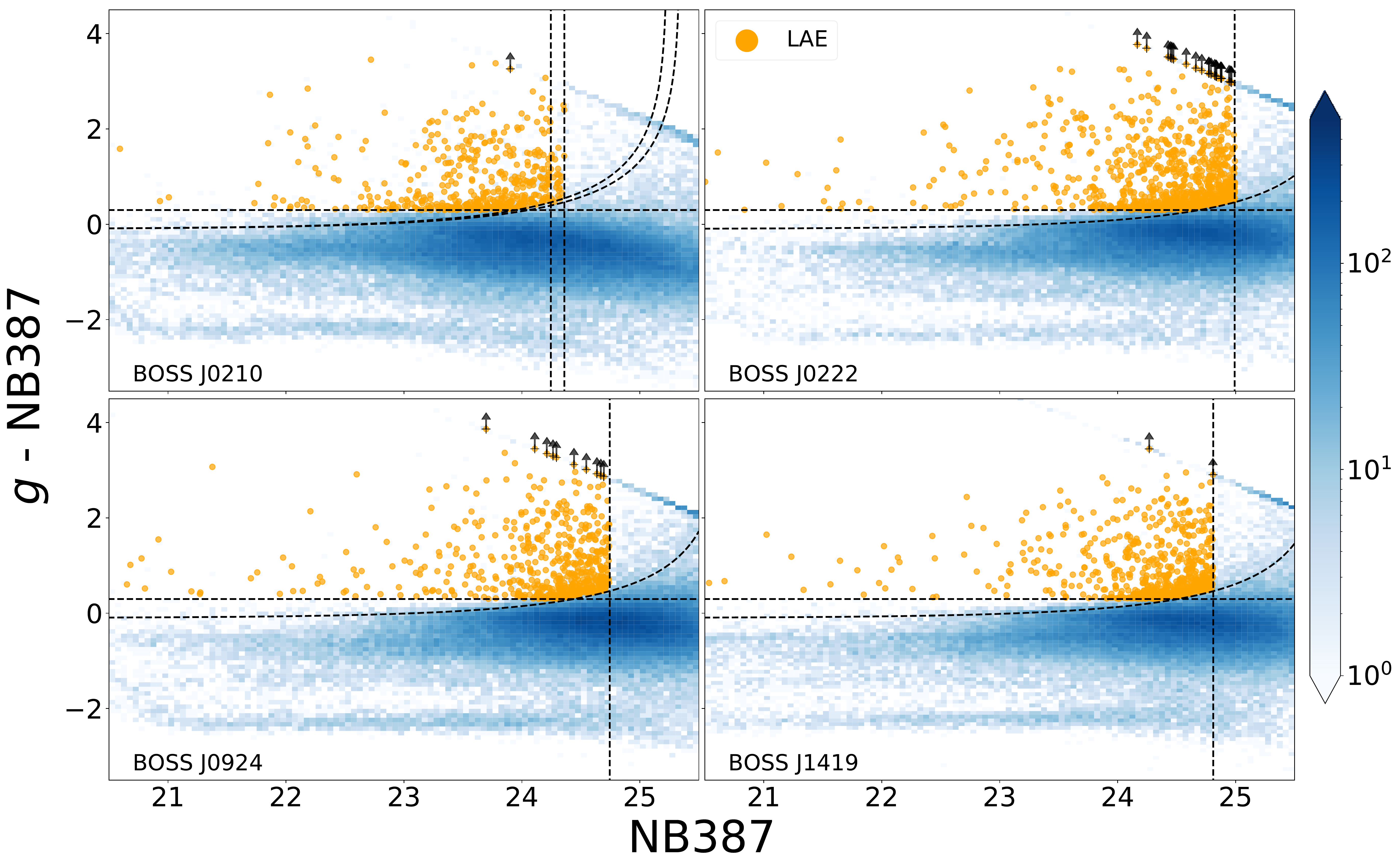}
    \caption{$g - {\rm NB387}$ vs. NB387 diagram for LAE selection in each field. 
    The two dimensional histogram bins the all detections from {\it SExtractor} and the bluer bins contain more objects. 
    The yellow points are the selected LAE candidates after visual inspection. 
    The three selection criteria are shown as the black dotted lines. 
    Specially, for the field $J0210$, the data is reduced in two versions of the {\it hscPipe} and there is a slight difference of the final image depth, so the criteria are overplotted for clarification.
    The black arrows indicate the LAE candidates with the $g$-band fainter than the respective $2\sigma$ limiting magnitude of each image, and the $g - {\rm NB387}$ shown for these objects are the lower limits.}
    \label{fig:lae_select}
\end{figure}

\begin{figure}[htb]
    \centering
    \includegraphics[width=\linewidth]{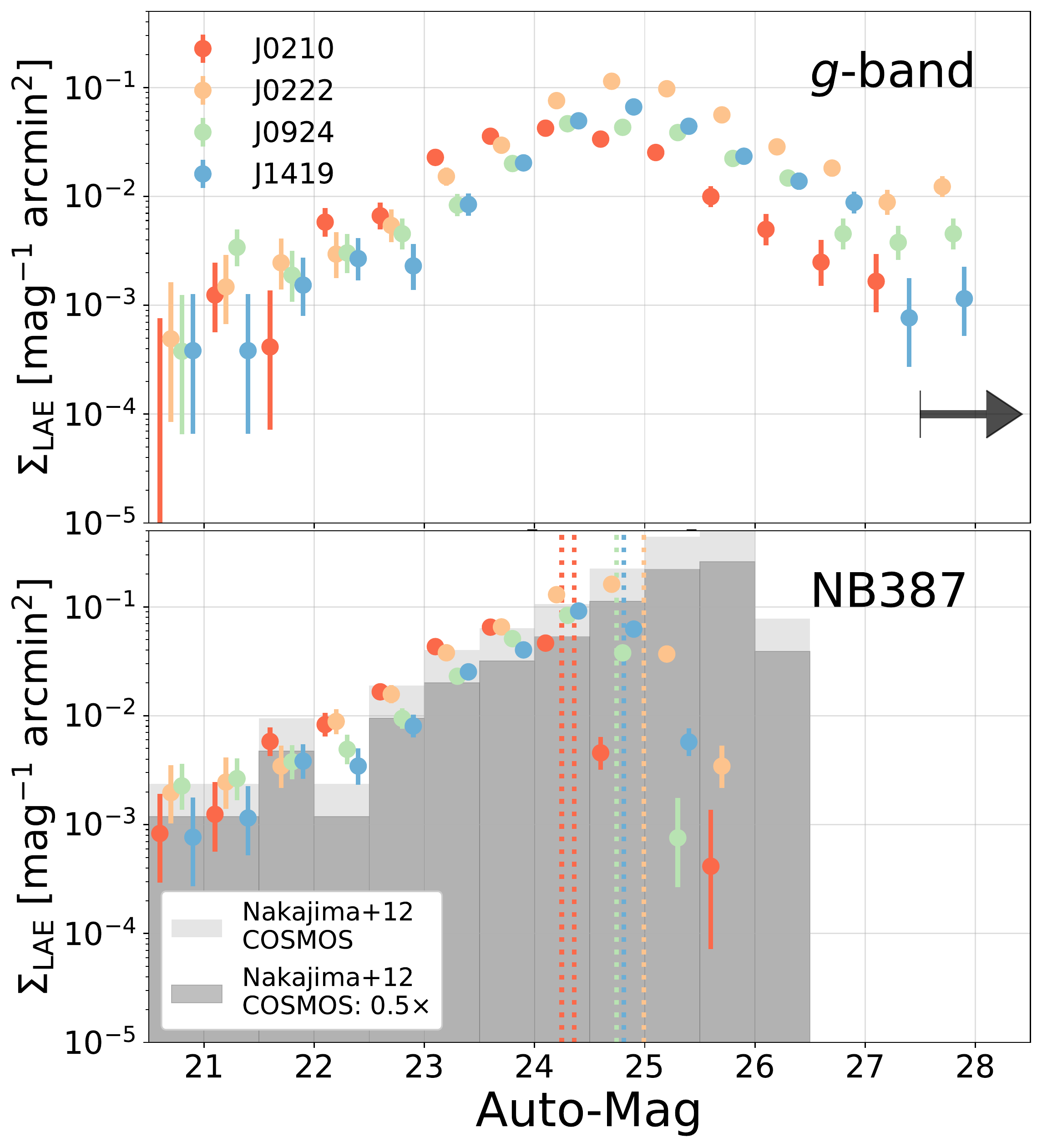}
    \caption{The Auto-Mag surface density distributions of the LAE candidates. 
    Red, yellow, green and blue histograms represent for the LAEs in $J0210$, $J0222$, $J0924$ and $J1419$ respectively. 
    Errorbars indicate the Poisson errors.
    {\it Upper:} $g$-band magnitudes. 
    The black arrow indicates that the rightmost bins include the faintest objects whose magnitudes are larger than the $2\sigma$ limiting mags.
    {\it Lower:} NB387 magnitudes. 
    The $5\sigma$ limiting aperture mags are indicated by the vertical dotted lines in respective colors.
    The $z\sim2.2$ LAEs in COSMOS selected by the Subaru/Suprime-Cam NB387 \citep{Nakajima+2012}, are also plotted for comparison as the light shade histogram.
    We also scale the number by a factor of 0.5 to roughly match the survey volume, and show it as the deep shade histogram.}
    \label{fig:lae_gmag}
\end{figure}

\subsection{\lya~Emitters Selection}
\label{sec:lae_select}
We use the color excess of the narrowband to the broadband as our LAE selection criteria, which is widely used in previous works \citep{Guaita+2010, Mawatari+2012, Nakajima+2012, Konno+2016, Zheng+2016}.
Though we have only the broadband data from the $g$-band on HSC for estimating the continuum, we prove here it is sufficient enough for the $z=2.18$ LAE selection.

In order to define the selection criteria,
we assume the LAE spectrum model at $z=2.0-2.5$ has a simple power law $f_{\lambda} = \lambda^{\beta}$ continuum and a correspondingly redshifted \lya~emission with a Gaussian-like profile, whose rest equivalent width $EW_0 = 20$ \AA. 
The IGM absorption is taken into account when we calculate the observed magnitude in each filter \citep{Inoue+2014}.
In addition to the $g$-band on Subaru/HSC, we include the adjacent broadband filters, the $u$-band on CFHT/MegaCam and the $r$-band on Subaru/HSC, to see the redshift evolution on the two color diagrams.
The tracks are shown in the Figure \ref{fig:lae_select_criteria}. 
The black curves indicate the color tracks of $g - {\rm NB387}$ vs. $u - {\rm NB387}$ in the left panel, and $g - {\rm NB387}$ vs. $r - {\rm NB387}$ in the right one.
Three different UV slopes $\beta$s, 0, -1.6 and -3.0, are shown in the both figures.

Meanwhile, we also overplot the predicted tracks of the possible contaminants, such as elliptical galaxies (age of 2, 5 and 13 Gyr denoted as Ell2, Ell5 and Ell13), starburst galaxies (M82 and N6090) and spiral galaxies (S0, Sa, Sb, Sc, Sd and Sdm) from the SWIRE library \citep{Polletta+2007} from redshift 0 to 3.0.
The homogeneously archived faint stars from SDSS used in Section \ref{sec:reduction} are also plotted.

From the color tracks, we find that the $r$-band is hardly helpful for the LAE selection, while the $u$-band may help to recover the extremely red ($\beta \sim 0$) populations.
However, given that the typical UV slope of the $z \sim 2$ LAE is found to be $\beta \sim -1.6$ \citep{Kusakabe+2019, Santos+2020}, we conclude that only the $g$-band is sufficient enough for our $z \sim 2.18$ LAE selection and a reasonable threshold of the color excess is $g - {\rm NB387} > 0.3$ to exclude most of the contaminants. 

To make the selection more confident in photometry, the color criteria are further defined as
\begin{equation}
    \begin{split}
    20.5~<~&\text{NB387}~\lesssim{\rm NB_{lim, 5\sigma}}, \\
    g -~&\text{NB387}~> 0.3, \\
    g -~&\text{NB387}~> 2\sigma(\text{NB387}) - 0.1.
    \end{split}
    \label{eq:lae_selection}
\end{equation}
The lower limit of the NB387 mag, 20.5, is set to avoid saturations, while the upper limit applies the 5$\sigma$ limiting magnitude to promise the reliability of NB387 detection. 
This upper limit for the field $J0924$ with a moderate depth, 24.74, corresponds to $1.94 \times 10^{42}\ \text{erg}\ \text{s}^{-1}$, which is $0.37 \times L^{*}_{\rm Ly\alpha}$ and the characteristic luminosity $L_{\text{Ly}\alpha}^* = 5.3 \times 10^{42}\ \text{erg}\ \text{s}^{-1}$ \citep{Konno+2016}.
The definition of the color error follows \citet{Shibuya+2018}:
\begin{equation}
    2\sigma(\text{NB387}) = -2.5~\text{log}_{10} \left( 1 - 2 \times \frac{\sqrt{f_{1\sigma,~\text{NB387}}^2 + f_{1\sigma,~g}^2 }}{f_{\text{NB387}}}\right).
    \label{eq:color_err}
\end{equation}
where the $2 \sigma$ follows a proper choice used in \citet{Nakajima+2012}. 
It aims to reject the false selection of the faint objects that pass the criteria due to statistical fluctuation around the $g-{\rm NB387}$ = -0.1, where the high-$z$ galaxy sequence lies on as described in Appendix \ref{appendix:zp_correct}.

The selected object passing the criteria are naturally filtered by the spatial masks, as the original object detection is performed with the masks applied.
Finally, we perform the visual check for each candidate to reject fake detections, like the hot pixels in the NB387 or the saturated pixels in the $g$-band image. 
We also check the cross-matches between our selected objects and the SDSS/(e)BOSS quasars at $z<2.15$ to discard the low-$z$ contaminants.
Eight are found in $J0210$, while six in $J0222$ and none in $J0924$ or $J1419$.
These known low-z quasars are removed from our LAE sample.
As a result, there are 465, 956, 585 and 636 LAE candidates selected out in the field $J0210$, $J0222$, $J0924$ and $J1419$ respectively, i.e., 2642 in total for the all four fields covering an effective area of 5.39 deg$^2$.
There are 4, 3, 0 and 1 proximity quasars from the SDSS/(e)BOSS matched to these LAE candidates in each field.
Specially, in the $J0210$ central region where 11 quasars reside in, three of the quasars are selected out as LAEs in our sample, while most others are too bright in the NB387 images and break the selection criterion NB387 $>20.5$.

As shown in Figure \ref{fig:lae_select}, the selected LAE candidates in the final catalog are plotted as the yellow points in the $g - {\rm NB387}$ vs. NB387 diagram, in which the all detections are binned in the two dimensional histogram coded by the blue color\footnote{A sequence appearing around $g-{\rm NB387} \sim -2.5$ is likely the stellar locus consisting of K and M-type stars, as suggested by the stellar locus in $ugr$ diagram \citep{Smolcic+2004}, and this is also supported by our random checks in matched SDSS spectra.}.

We show the $g$-band and NB387 magnitude distributions in surface density of the LAE candidates in Figure \ref{fig:lae_gmag}, with the Poisson errors estimated by the statistics proposed in \citet{Gehrels1986}.
In both filters, the J0210 and J0222 are found with the excess at around 23--24 mags.
The Auto-Mags are shown here for the fair comparison of total magnitudes with literature, and we overplot the $z\sim2.2$ LAE sample in Cosmic Evolution Survey (COSMOS) field from \citet{Nakajima+2012} that are selected by the Subaru/Suprime-Cam NB387 ($\lambda_0=3870$ \AA, FWHM $=94$ \AA).
As their FWHM is almost twice to the HSC NB387, corresponding to a roughly double survey volume, we also show the case scaled with a factor of 0.5.
The distribution shapes are almost consistent, but all of our four fields show out number excesses up around the limiting depth compared to the scaled numbers in COSMOS, although the excesses in $J0924$ and $J1419$ are not as significant as $J0210$ and $J0222$.
The excess is not surprising as our fields are selected to contain potential overdensities.

Comparing with other galaxy--IGM correlation studies, we note that while LAEs are expected to be younger and less massive than the more mature LBGs in Keck Baryonic Structure Survey \citep[KBSS,][]{Rudie+2012, Chen+2020} and $Ks$-selected photo-$z$ galaxies \citep{Mukae+2017}, our samples also reach deeper regarding the UV continuum given the depth limit of $R\sim 25.5$ in KBSS and $g \sim 26.4$ in \citet{Mukae+2017}.

\subsection{Potential Contaminants}
\label{sec:imcomple}
Besides the LAEs at $z\sim2.2$, some of the lower-$z$ emitters may also pass our selection criteria.
For the filter NB387, the contaminants are mainly considered as {\sc [Oii]} emitters at $z=0.036\pm0.008$. 
But, the survey volume at such redshift range is much smaller than that at the $z\sim2.2$, and the ratio reaches $0.2\%$.
Given the low-$z$ {\sc [Oii]} emitters luminosity function from \citet{Ciardullo+2013} and our NB387 image depth, we can estimate that the detected number is $\sim0.05$ in one HSC FoV.
We conclude that the contamination rate of low-z {\sc [Oii]} emitters is negligible in our sample.
In addition, {\sc Ciii]} $\lambda1909$ at $z\sim1$ and {\sc Civ} $\lambda1548$ at $z\sim1.5$ can be also the interlopers.
However, according to \citet{Konno+2016}, these emitters should be probable AGNs, as the objects passing our selection criteria yield the $EW_0 \gtrsim 30$ \AA, which is much larger than that in the typical star-forming galaxies. 

In the literature working on fields like SXDS, COSMOS, HDFN, SSA22 and E-CDFS fields \citep{Guaita+2010, Zheng+2016, Konno+2016}, they use the detection in the database covering multi-wavelengths, e.g., the X-ray, UV and radio, to exclude the low-$z$ AGN contaminants.
In our case, however, we search for overdense fields from the whole (e)BOSS survey, and therefore, the deep multi-wavelength data is not available for testing the AGNs in this work.
Instead, we refer to the literature aforementioned and find that the contamination rate of the LAE selections at $z\sim2.2$ is commonly $\sim 10-15\%$, and \citet{Sobral+2017} also confirm this number spectroscopically.

We test this contamination estimate for the case of HSC/NB387 by utilizing the COSMOS data, the NB387 data from the Cosmic HydrOen Reionization Unveiled with Subaru (CHORUS; Inoue et al. 2020, submitted) survey and the DEIMOS 10K spectroscopic survey catalog \citep{Hasinger+2018}. 
It yields a contamination rate of $\sim 15\%$ in our LAE selection, and about 2/3 of the interlopers are likely to be the {\sc Civ} emitters at $z \sim 1.5$, which shows a good consistency to what have been stated in the previous studies.
As this contamination level is not crucial to our statistical study,
we keep all the selected LAE candidates in our overdensity maps as well as the correlation analysis performed in the following sections.

\section{Results} 
\label{sec:result}

\subsection{LAE Overdensity Map}
\label{sec:overdensity}
\begin{figure*}[hbt!]
    \centering
    \includegraphics[width=\textwidth]{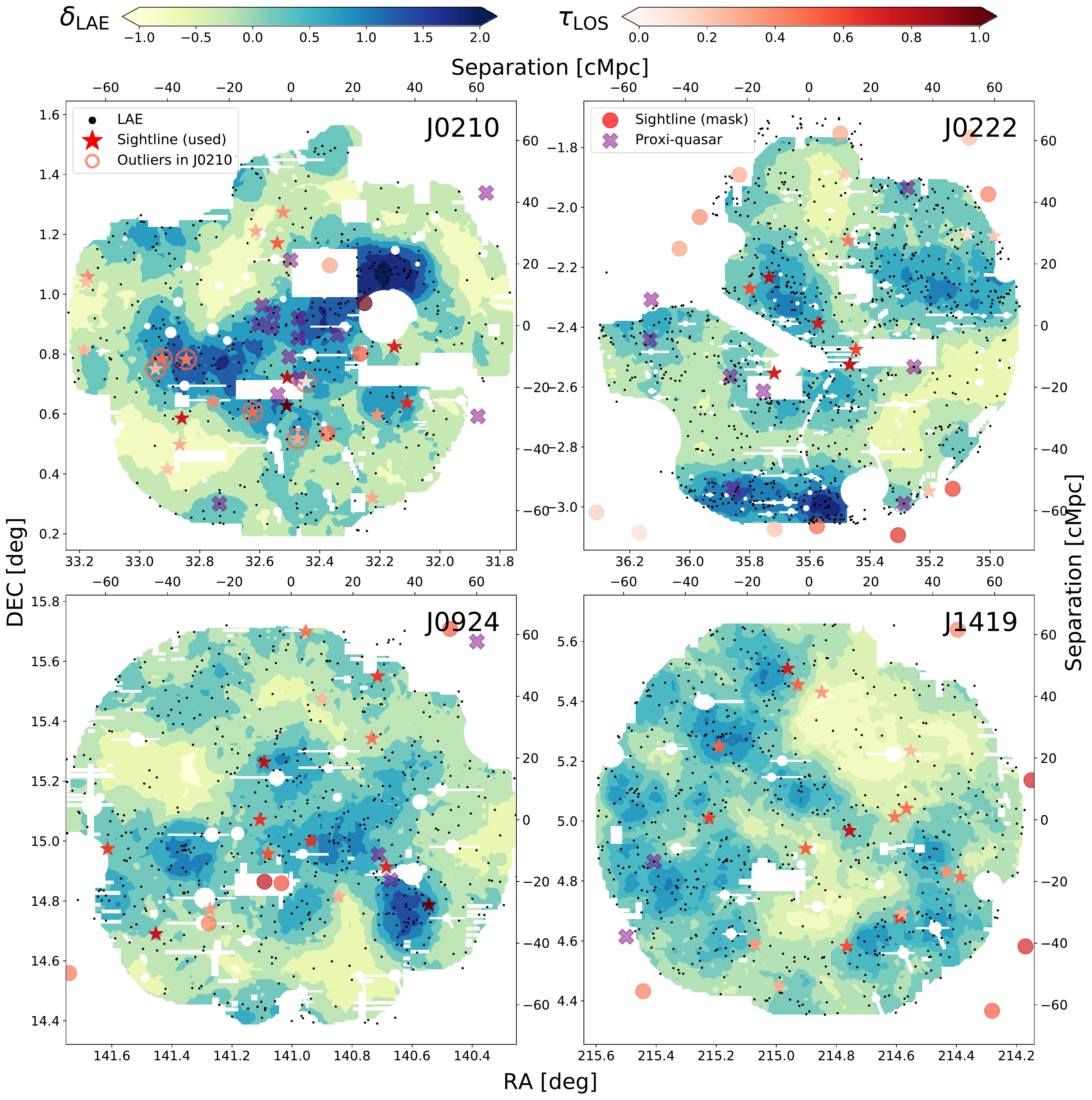}
    \caption{Overdensity maps for the four selected fields $J0210$, $J0222$, $J0924$ and $J1419$. 
    Black points represent the LAE candidates. 
    The blue contour in the background is scaled by the LAE overdensity $\delta_{\rm LAE}$ on a scale of $r=10$ cMpc.
    The red stars and dots are the positions of used LoSs and masked LoSs respectively, with the color coded by effective optical depth on a scale of 15 $h^{-1}$ cMpc.
    The purple crosses represent the proximity (e)BOSS quasars with $2.15<z<2.20$.
    White regions are the masks used to exclude regions with low S/N signals, saturation around bright stars or serious stray light.
    The circles highlight the outliers in $J0210$, and details can be checked in the text in Section \ref{fig:od2_direct}.}
    \label{fig:lae_od}
\end{figure*}

The sky distribution of the selected LAE candidates is shown in the Figure \ref{fig:lae_od}.
We calculate the galaxy overdensity over each field to construct the overdensity maps.
The overdensity is defined as 
\begin{equation}
    \delta_{{\rm LAE}} = \frac{{\rm N}_{i,\ {\rm LAE}} - \left< {\rm N}_{\rm LAE} \right> }{\left< {\rm N}_{\rm LAE} \right>},
    \label{eq:od_def}
\end{equation}
where the ${\rm N}_{i,\ {\rm LAE}}$ is the number of galaxies counted within an aperture with the fixed radius, and the $\left< {\rm N}_{\rm LAE} \right>$ is the mean number of galaxies in an aperture averaged over each field respectively. 

The aperture size is set as 10 cMpc ($\sim 6$\arcmin) in radius, which yields a mean number of LAEs counted in an aperture $> 10$, and so we can have a mean S/N $>$ 3 signal assuming the Poisson statistics for counting.
The map is constructed through a 128 $\times$ 128 meshgrid for each field, which corresponds to a resolution of $\sim$ 1 cMpc.
When calculating the mean number density, we exclude the apertures covering the masked regions for more than 10\%, a strict criterion to keep the mean number estimate robust. 
While drawing the overdensity map, we exclude the apertures that are masked out over 50\%, a relaxed criterion to show more accepted regions.
The mean number $\left<N_{\rm LAE}\right>$ (standard deviation $\sigma_{\rm LAE}$) within a radius $r=10$ cMpc aperture for $J0210$, $J0222$, $J0924$ and $1419$ are 10.7 (6.0), 23.5 (10.1), 12.6 (4.9) and 14.0 (5.4) respectively.
The smaller mean number in $J0210$ and larger number in $J0222$ are mainly originated from the image depth difference.

In Figure \ref{fig:lae_od}, the blue contours in the background show the overdensity.
Masked regions that are defined in Section \ref{sec:photometry} are also shown as the white areas.
The LAEs and the proximity quasars with $2.15<z<2.20$ checked when selecting candidate fields are both shown for each field.
The position of LoSs are also marked as the red stars with the color coded by the effective optical depth \optdep.

\subsection{Notes on Individual Fields}
\label{sec:note_field}
More quantitative discussions on the overdensity catalog will be presented in Cai et al. (in prep.), and we just have a brief overlook here.

In the Figure \ref{fig:lae_od}, we find a large filamentary structure at the center of the field $J0210$ as well as the structures with weaker significance, which are likely to be the sheet-like structures, around the nodes at the ends.
The field is traced by both the central grouping LoSs with strong \lya~absorption and a group of quasars clustering within an area of $\sim(40~{\rm cMpc})^{2}$ at $2.15<z<2.20$.
The filamentary structure they are associated with extends for about $100$ cMpc, and the peak density of one node with \overden$\sim3$ reaches the significance of over $6\sigma$.
This result supports that the combination of using both tracers seems to effectively hint the unique LSS, as also suggested in \citet{Cai+2017a}. 
Given what have reported in the previous studies on the correlation between multiple quasar environments and the galaxy overdensity \citep{Hennawi+2015, Cai+2017b, Onoue+2018, Mukae+2019}, the emergence of the grouping quasars suggests the filament is much different from the typical environments at $z\sim2$, and $J0210$ will be considered individually in the following parts.
The uniqueness of the structure in $J0210$ is out of the scope of this paper, and further discussion will be made in our future paper.

As to the $J0222$, one can find that this field is seriously affected by bright stars in and around the FoV, which results in large masked areas with strange patterns.
A weak clump with an overdensity $\delta_{\rm LAE} \sim 1.0$ over a 20 cMpc length scale is found close to the central region, likely associated with the central group of high \optdep~LoSs. 
Another clump with comparable significance appears at the west side, but it seems to be independent from the central structure.
Interestingly, a large filamentary structure with an overdense peak $\delta_{\rm LAE} > 1.6$ appears at the southern boundary of the FoV.
There are nearby LoSs in the vicinity with relatively high \optdep, but they are out of our pointing FoV.
So, this structure is not found intendedly by the strong IGM \lya~absorption, but just by chance.

In the field $J0924$, we mainly use the central four grouping LoSs with high \optdep~as tracers.
But in the central area, we do not find a structure with significant overdensity based on this LAE sample.
Within the $J0924$, several peaks have moderate overdensities $\delta_{\rm LAE} > 0.8$ that are comparable with or surpass the central structure. 
The most overdense structure is found at the southwest of the field, which is close to two LoSs with strong \lya~absorption.
The peak of the structure has an overdensity measured over 1.2, and it extends for about 30 cMpc.

The field $J1419$ shows more structures in the clumpy shapes.
Although there are four LoSs with \optdep$>0.6$, they are more scattered with distances of $\sim40-100$ cMpc to each others compared to those in other fields.
Hence the coherently strong absorption is expected to be less significant.
But instead, the number of LoSs in this field is appreciable for the correlation analysis.
Five peaks with the moderate \overden~$>0.6$ can be found in various regions, but no extreme overdense or extended structure is in this field.
On the contrary, a large void with a size of $\sim 50\times60$ cMpc$^2$ emerges at the northwest of the FoV.

\subsection{Correlation between Galaxy and IGM {\sc Hi}}
\label{sec:od2_direct}
The past observational studies on a large scale correlation are still restricted by the limits in both the FoV and the depth.
We have described some relations between the LoSs and the overdensities qualitatively in Section \ref{sec:overdensity}, 
and from this section, we will have more quantitative analysis on such correlation in statistics.

\begin{figure*}[hbt!]
    \centering
    \includegraphics[width=0.8\textwidth]{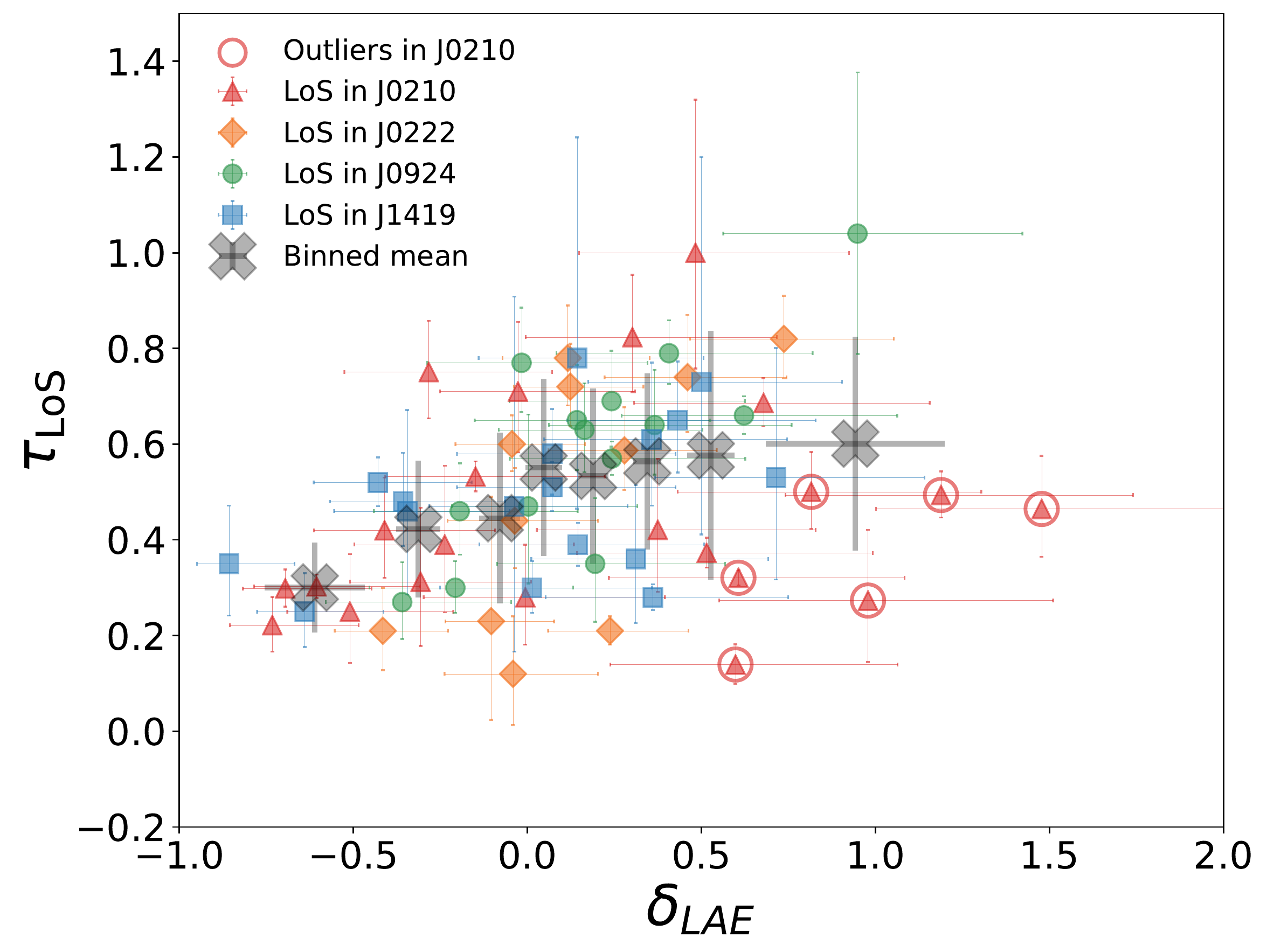}
    \caption{Correlation between LAE overdensity \overden~and effective optical depth \optdep~at the positions of LoS. Red triangle, yellow diamond, green circle and blue square represent the original data points in fields $J0210$, $J0222$, $J0924$ and $J1419$ respectively. 
    The number of the LoSs are 63 for 4 fields on a scale of 10 cMpc with less than 50\% masked vicinity. 
    The grey corsses are the binned data points with  the 1$\sigma$ standard deviation.
    Except for the most overdense bin which is dominated by the data points in $J0210$, a clear increasing trend shows out.
    The outliers in $J0210$ with close spatial distribution are highlighted by red circles.}
    \label{fig:od2_direct}
\end{figure*}

To quantify the correlation, we calculate the overdensity on the scale of 10 cMpc in radius, at the positions of the clean LoSs. 
Similar to the Section \ref{sec:overdensity}, we discard the LoSs whose vicinity are masked out by more than 50\%, but as the result, no LoS is removed in this process and the number of remaining LoSs is still 64.
We assume the density in the masked regions to be the mean value in each field respectively.
Errors are estimated as the Poisson noise using the statistics proposed in \citet{Gehrels1986}, which is the dominant uncertainty due to the small number statistics \citep{Cai+2017a}.
Then we can compare the LAE overdensity \overden~and the effective optical depth \optdep~measured for the LoSs, whose error is derived from the error of mean flux in pixel statistics, to investigate the correlation.
Figure \ref{fig:od2_direct} shows the result.
As we can see from the figure, the error for \overden~suffers from the Poisson statistics with a small number of shots (LAEs) in each measured aperture.
While for \optdep, the large error is mainly due to the relatively low S/N of the quasar spectra at the NB387 sensitive wavelength, which is close to the blue-end of the response range of the SDSS spectrograph. 
Note that we have discarded LoSs with continuum-to-noise ratio smaller than 2. 

A tentative positive correlation can be found intuitively in the figure albeit, though with a large scatter.
We perform the Spearman's rank correlation test for the full data sample, and the result shows the Spearman's rank correlation coefficient as $\rho_S = 0.384^{+0.015}_{-0.037}$ with a $P$-value $=0.09\%$. 
The uncertainty of $\rho_S$ is estimated by performing a Monte Carlo simulation by fluctuating the data points within their errors. 
We make 10,000 runs to pull sets of pseudo data from the Gaussian distributions, whose mean $\mu$ and standard deviation $\sigma$ are the observed data and the corresponding error.
The shown values are the 16\%, 50\% and 84\% rank of the simulated $\rho_S$ results.
It proves a moderately positive correlation with strong confidence between the LAE overdensity and IGM effective optical depth, based on the LoSs that are randomly distributed on the areas extended to scales over 100 cMpc at $z>2$.

We find that the large scatter in Figure \ref{fig:od2_direct} might be largely contributed by the LoSs in $J0210$ (red triangles), which contains a unique structure and has a shallower limiting magnitude.
If we exclude $J0210$, the Spearman's rank correlation increases largely to an $\rho_S$ = $0.541^{+0.037}_{-0.051}$ with a $P$-value $<0.01\%$.
The NB387 limiting magnitude of $J0210$ is shallower than others and the selected LAEs distribute at the relatively bright-end.
The bright-end LAEs can result in the overestimated overdensities compared to other fields \citep{Lee+2014b, Casey+2015}.
In this case, the bias from the potentially different spatial distributions of bright and faint galaxies can enlarge the scatter of overdensity.
We perform the same correlation analysis by limiting the LAE NB387 magnitude to 24.3 for all four fields, and there are 451, 288, 264 and 248 LAEs left in $J0210$, $J0222$, $J0924$ and $J1419$ respectively, but the results with ($\rho_S=0.388^{+0.026}_{-0.045}$) and without ($\rho_S=0.502^{+0.031}_{-0.061}$) $J0210$ are consistent with those shown previously within the uncertainty, and cannot explain the significant difference.
Therefore, the limit of the bright-end is unlikely to be the driven origin.

Alternatively, the difference of $\rho_{S}$ can also be originated from the field variation in the correlation.
The found large filament and the existence of the grouping proximity quasars indicate that the structures in $J0210$ are probably different from other fields.
More fields will be required for the more robust statistics in the future.

The binning data\footnote{The bins are made by sorting data points according to their \overden~and splitting the nearest eight LoSs into one bin.
There are eight bins for the 63 LoSs in total, with 7 LoSs in the largest \overden~bin, and the error is the 1$\sigma$ standard deviation at each bin.}
shows a clearer trend intuitively, which is overlaid as the grey crosses in Figure \ref{fig:od2_direct}. 
The \optdep~increases with the \overden~at all range, though interestingly, the pace of increasing seems to be slower and the trend becomes flatter when \overden~$\gtrsim0.2$.
We notice the trend at the overdense end is likely dominated by the $J0210$ LoSs contributing in the \overden~$>0.5$ bins.
Especially, some of these LoSs are spatially close, and we 
highlight these special LoSs, hereinafter referred to as {\it outliers}, by circling them out in Figure \ref{fig:od2_direct}, and their sky distributions are also shown in Figure \ref{fig:lae_od} with the same symbol.
We can find that the outliers cluster at two regions in $J0210$, which are close to the node of the filament. 
Considering that the $J0210$ LoSs do not show a large scatter at the smaller \overden~bins, the field variation, instead of the bright-end limit, is again favored to be the reason for the $\rho_S$ difference between the cases with and without $J0210$.

Therefore, it might indicate that different physical processes may have taken place in the $J0210$ filament compared to the typical structures at the same redshift. 
The lower \optdep~of the outliers than those of other LoSs can suggest either the lack of IGM {\sc Hi} in $J0210$ or the LAE deficit in other fields, or both.
As mentioned above, when limiting the LAE sample with NB387 magnitude up to 24.3, there are 451 LAEs in $J0210$.
The number is a factor of $>1.5$ larger than the cases in the other fields, suggesting the number excess in $J0210$.
Meanwhile in passing, we note that \citet{Momose+2020b} have found that the LAEs might be residing slightly off-centered from the most highest density regions identified by the {\sc Hi} tomography. 
This is suggestive and is consistent with the lower {\optdep} values in $J0210$, although a larger and deeper sample of LAEs and many more higher-resolution LoSs would be needed to say something more definitive. 

\subsection{Cross-correlation Analysis}
\label{sec:ccf_lae_los}
\begin{figure*}[hbt!]
    \centering
    \includegraphics[width=3.5in]{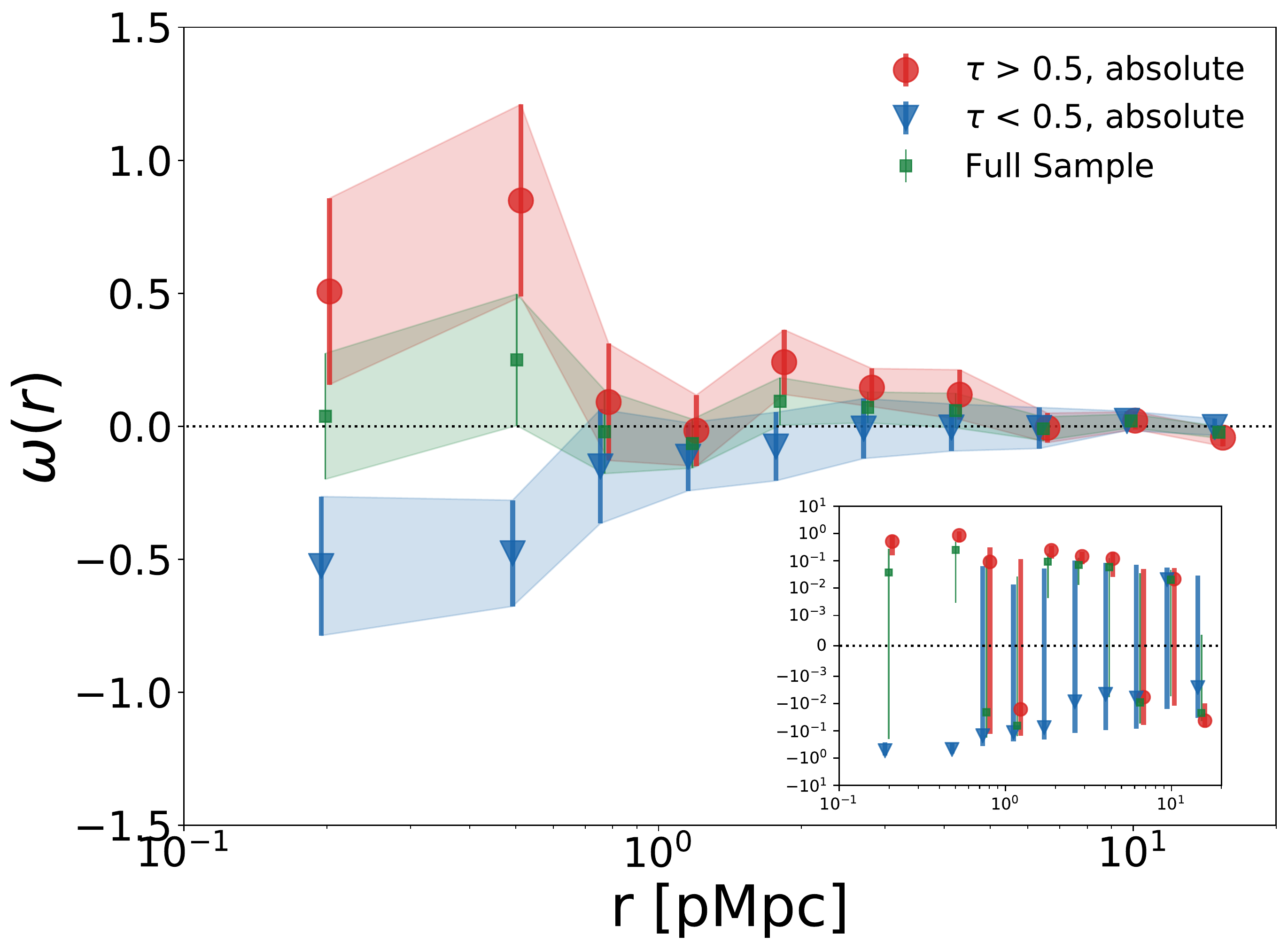}
    \includegraphics[width=3.5in]{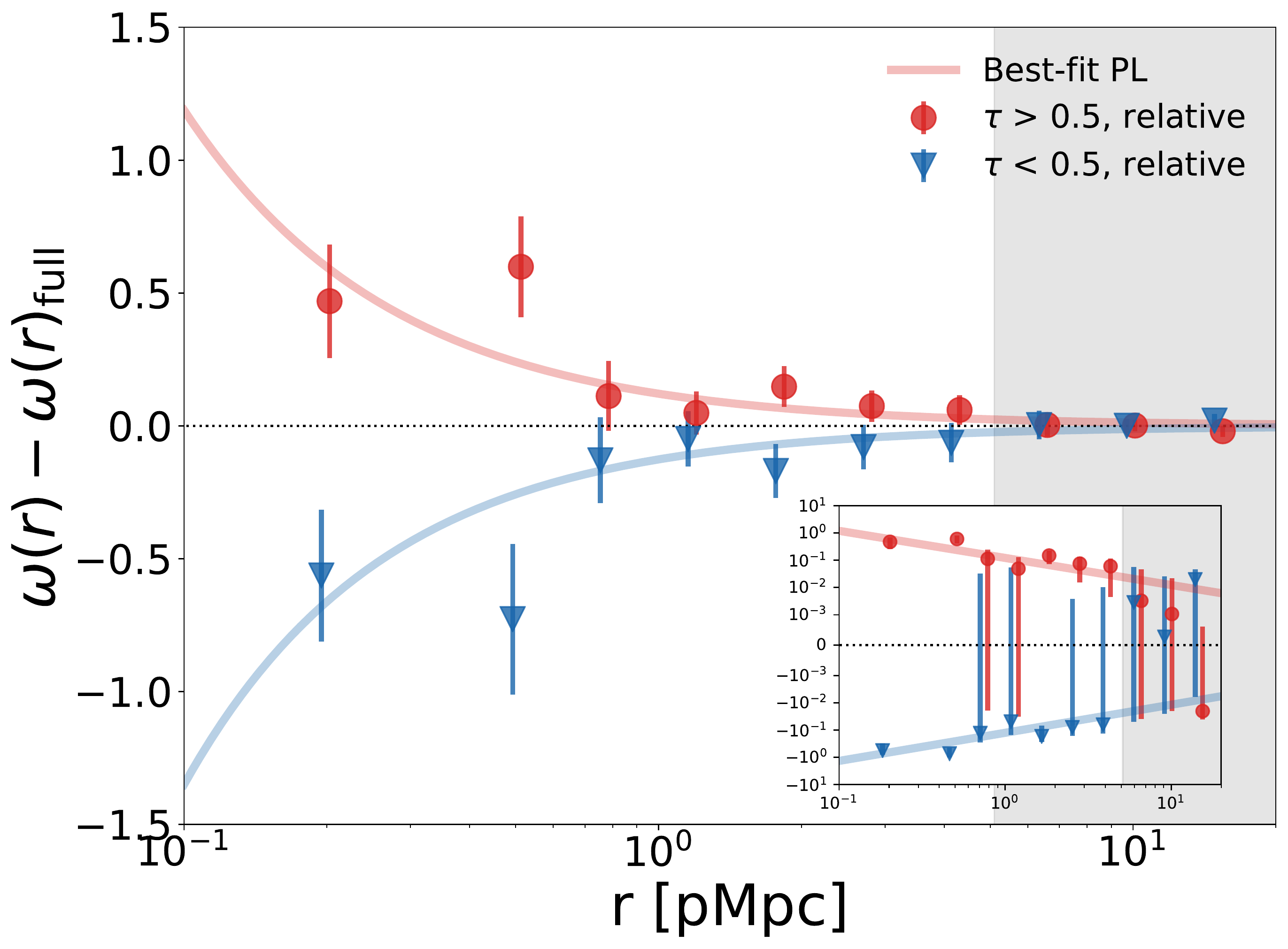}
    \caption{Cross-correlation function (CCF) between LAEs and LoSs for the high \optdep/low \optdep~subsamples. 
    Red points and curves are the \optdep~$>0.5$ subsample and the corresponding fit power law model, while blue points and curves are for the \optdep~$<0.5$ subsample.
    Data points for different subsamples at each bin are slight shifted along the $r$-axis for clarification.
    In both panels, the major figures are shown in linear scale while the inset figures are shown in log scale.
    {\it Left panel}: original CCFs are shown for the two subsamples in red/blue, and the full sample in green, whose LoSs consist of the two subsamples.
    Shaded regions are the uncertainties from Jackknife resampling.
    {\it Right panel}: relative CCFs are calculated by subtracting the full sample signal $\omega(r)_{\rm full}$ from the original CCF of each subsample $\omega(r)$.
    The solid curves are the best-fit power law models for all of the data points.
    The grey shaded region indicates the separation limit where the signal becomes noisy.}
    \label{fig:ccf_linear_log}
\end{figure*}
Along with the analysis based on the local overdensity of LoSs, a more general analysis can be made with the galaxy--IGM {\sc Hi} correlation.
Correlation of the spatial distribution can be translated as the clustering properties between the two populations of objects, and to quantify the clustering strength, the two-point cross-correlation function (CCF) can work as an ideal tool.

We divide the LoSs into two subsamples according to the measured \optdep. 
For the purpose of having a comparable number of LoSs in the two subsamples, we set the criterion as \optdep~= 0.5.
LoSs with \optdep~$>0.5$ are called as the high \optdep~LoSs, while LoSs with \optdep~$<0.5$ are correspondingly called as the low \optdep~LoSs.
In this case, if we use the full sample, then the number of LoSs for high/low \optdep~subsample is 30/34 respectively, and if we exclude field $J0210$, then the number of LoSs for high/low \optdep~subsample changes to 23/19.
We mainly discuss the case including the $J0210$ LoSs for CCFs.
Although there are outlying LoSs found in Section \ref{sec:od2_direct} in $J0210$,
we note that only 6 outliers are pinpointed while there are 64 LoSs in total.
The statistics like CCF is unlikely to be biased.

We use the angular CCF $\omega(\theta)$, or the so-called projected CCF $\omega({r})$ if the angular separation is translated into projected physical distance, for our analysis.
To estimate the $\omega(\theta)$, we apply the estimator proposed by \citet{LS93}, which can be better constrained in errors, to compare the data pairs against the randomly distributed points:
\begin{equation}
    \omega(\theta) 
    = \frac{ D_{\rm LAE}D_{\rm LoS}(\theta)
            - D_{\rm LAE}R(\theta) 
            - D_{\rm LoS}R(\theta)
            + RR(\theta)}
           {RR(\theta)},
    \label{eq:ccf_def}
\end{equation}
where $D_{\rm LAE}D_{\rm LoS}(\theta)$, $D_{\rm LAE}R(\theta)$, $D_{\rm LoS}R(\theta)$ and $RR(\theta)$ are the normalized LAE--LoS, LAE--Random, LoS--Random and Random--Random pairs counted at the separation of an angle $\theta$ within an interval of $\delta\theta$.
The normalization factor is the total pair number of each term.

To keep the statistics significant, we choose the right boundary of the innermost bin as 0.013 deg ($\sim 0.4$ pMpc at $z=2.2$) so that there are $>10$ pairs at the bin in one subsample, reaching S/N $>$ 3 in Poisson statistics. 
Ten bins are set for the calculation extending up to 0.6 deg\footnote{We test the following results by varying the bin size, and we confirm that our major results are not sensitive to the bin determination.}.
Note here that $D_{\rm LAE}D_{\rm LoS}(\theta)$ represents the LAE--LoS pairs, not the LAE--absorber pairs. 
We do not use the information of location along the LoS of the absorbers, because the exact LAE redshifts are unknown within $2.15<z<2.20$, and the LoS-direction distance is meaningless even if we know where the absorbers are. 
This is why we use the projected CCF, but not the 3D one.

The error for the CCF is estimated by the Jackknife resampling, which can also take the field fluctuation into account.
To do the resampling, we split each HSC field into $5 \times 5$ square sub-fields, and the sub-fields that are overlapped by over 50\% mask regions are excluded to ensure a sufficient number of pairs in each sub-field.
Following \citet{Norberg+2009}, we denote \textbf{\it i} as the calculating log scale bin, and \textbf{\it k} as the resampling run.
In the $k^{\rm th}$ run, we skip the $k^{\rm th}$ sub-field and perform an identical CCF calculation as for the full field sample.
Then the variance of the statistics of our interest, i.e., $\omega(r)$, can be derived for the ${i^{\rm th}}$ bin:
\begin{equation}
    \sigma_i = \frac{N_{\rm sub} - 1}{N_{\rm sub}} \sum^{N_{\rm sub}}_{k=1} \left( \omega_{i, k} - \overline{\omega_i}\right)^2,
    \label{eq:ccf_jr_var}
\end{equation}
where the $\overline{\omega_i}$ is the mean over all resampling runs given by $\overline{\omega_i} = \sum^{N_{\rm sub}}_{k=1}\frac{\omega_{i, k}}{N_{\rm sub}}$ at the $i^{\rm th}$ bin.

As we described above, the projected CCF does not rely on the information of LoS-direction location.
Without considering the \lya~absorption, the LoSs should be viewed as being selected homogeneously from the sky and they are not dependent on the foreground IGM at $z\sim2.2$. 
Therefore, if the LoS number is infinite, a full sample without being split by the \optdep~is expected to have a null CCF signal.

However, our sample size is limited in fact, and this may involve an artificial signal into the CCF. 
We firstly check the CCF for the full sample combining the high \optdep~and low \optdep~LoSs, and the result is shown as the green points in the left panel of Figure \ref{fig:ccf_linear_log}. 
Although the full sample has much weaker signal than any subsample, which is clearer in linear scale by comparing the green points with blue/red points, they do not exactly equal to zero. 
This effect is due to a limited sample size. 
For the purpose of the clearer comparison, we subtract the amplitude of the full sample CCF $\omega(r)_{\rm full}$ from that of each subsample CCF $\omega(r)$, and we call the reduced signal as the relative CCF, i.e., $\omega(r) - \omega(r)_{\rm full}$, which is shown in the right panel of Figure \ref{fig:ccf_linear_log}.
Data of subsamples at each bin is slightly shifted along the $r$-axis in the figures for clarification.

From the both panels in Figure \ref{fig:ccf_linear_log}, we find the high \optdep~subsample shows a continuous positive signal from the innermost bins up to a separation $r \approx 4$ proper-Mpc (pMpc). 
On the contrary, the low \optdep~CCF stays negative in the same distance range.
By varying the bin size, this characteristic distance changes by smaller than 1 pMpc.
This result suggests that up to a scale of $4\pm1$ pMpc ($\sim 13\pm3$~cMpc at $z=2.2$), LAEs tend to cluster in the regions rich in gas, indicated by the high \optdep~LoS, and avoid the low \optdep~region where the gas is less abundant\footnote{As the accurate LAE redshifts are unknown, one can view the \optdep~estimated at the absorption spike works as the upper limit constraining the intrinsic {\sc Hi} associated with the LAEs around a LoS. 
This is why the $\tau_{\rm LoS}<0.5$ subsample can show a negative CCF, even though many of them still own the $\tau_{\rm LoS}$ higher than the cosmic mean value.}. 
We also notice the two bins at $\sim 0.8-1.0$ pMpc tends to be consistent with zero, suggesting a weak signal at the distance.

Interestingly, the CCF shown in the right panel can be well fitted by a power law:
\begin{equation}
    \omega(r) = \pm \left(\frac{r}{r_0} \right)^{-\gamma},
    \label{eq:ccf_powerlaw}
\end{equation}
where the $r_0$ is called as clustering length that makes $\omega(r_0)=1$, and it can be an indicator of the clustering strength.
We fit the binned data points with the power law by using least-square method with Levenberg-Marquardt algorithm, and the fitting curves are shown in corresponding colors in the Figure \ref{fig:ccf_linear_log}. 
The best fit parameters ($\gamma$, $r_0$) with the errors estimated from the 10,000 Monte Carlo perturbed simulations, similar to Section \ref{sec:od2_direct}, are ($0.99^{+0.54}_{-0.17}$, $0.12^{+0.05}_{-0.03}$ pMpc) and ($1.03^{+0.83}_{-0.21}$, $0.13^{+0.06}_{-0.02}$ pMpc) for the high and low \optdep~subsamples respectively, and they are also summarized in Table \ref{tab:ccf_param} in Appendix \ref{appendix:ccf_subsample}.  
The $r_0$ for both subsamples are of an order of $0.1$\,pMpc, which is much smaller than the typical clustering strength in the case of galaxy-galaxy clustering derived from 3D CCF, i.e., several pMpc. 
This indicates that the strength of the LAE$-$IGM {\sc Hi} clustering is not very strong, thought it is still significant enough for being detected based on our samples for the projected CCF.
We note that \citet{Momose+2020b} obtained somewhat stronger 3D CCF signal between LAEs and CLAMATO {\sc Hi} absorption data with $r_0 = 0.78\,h^{-1}$\,cMpc, which corresponds to $\sim 0.35$ pMpc for $z\sim 2.2$ in our cosmology.

We test whether the results will be changed, if we exclude the field $J0210$, or if we change the \optdep$=0.5$ criterion to separate the LoSs into subsamples.
We do not find that such factors have significant impacts on our results, and details can be found in Appendix \ref{appendix:ccf_subsample}.

\subsection{Average Optical Depth Profile to LAEs}
\label{sec:average_tau}

We can further trace down to the circumgalactic medium (CGM) scale using  our LAE and LoS samples.
The aforementioned analyses mainly focus on $r \gtrsim 1$ cMpc.
Because the overdensity-based analysis requires a large enough aperture to overcome the small number statistics when counting galaxies.
In the CCF analysis, we need to divide LoSs into high/low \optdep~subsamples, which makes a drop in the sample size by at least a factor of two.
This can be extremely problematic for the smallest separation bin, which pushes us to set the innermost bin as large as 0.013 deg.

For the purpose of studying a smaller scale down to sub-cMpc, or $\sim200$ kpc in physical length, where CGM is supposed to surround the host galaxies, we perform another analysis that is similar to the stacking technique in concept.
We derive the average radial distribution of the IGM optical depth that is averaged over all LoSs within a ring-like bin centered at a specific LAE.
We then further calculate a mean over all LAEs, named as the average \optdep~profile $\left<\tau\right>$, where $d$ is the distance from the stacked LAE.
To emphasize the excess level, we define the fluctuation of the
$\left<\tau\right>$ as 
\begin{equation}
    \delta_{\left<\tau\right>}(d) 
    = \frac{\left<\tau\right> - \left< \tau \right>_{\rm tot}} 
           {\left< \tau \right>_{\rm tot}},
    \label{eq:avetau_fluc}
\end{equation}
where $\left<\tau\right>_{\rm tot}$ is the mean over the radial direction.
We first calculate the $\left<\tau\right>_{\rm tot}$ over a large distance range $0 < d < 0.3 $ deg, or $ 0 < 0 < 9.2$ pMpc at $z=2.2$, in the two cases, i.e., the coarse bin with a spatial resolution of $\Delta d = 600$ pkpc and the finer bin with the a higher resolution $\Delta d = 200$ pkpc \footnote{The two cases are chosen because: (1) they are concerned in physics as the $200$ pkpc is a typical scale of CGM and $600$ pkpc is persuasively far enough to be in the IGM regime; (2) signals only exceed the 84\% ranks in random LoSs with these two choices to draw a meaningful result after testing various bin sizes.}.
Then the $\left<\tau\right>$ can be derived based on the bins, and the signal of CGM is expected in the inner regions.
\citet{Momose+2020b} also found CCF signal at CGM scales between LAEs and {\sc Hi} absorption with an interesting plateau in the central few hundreds pkpc (see their Fig.\,9).
\begin{figure}[htb!]
    \centering
    \includegraphics[width=\linewidth]{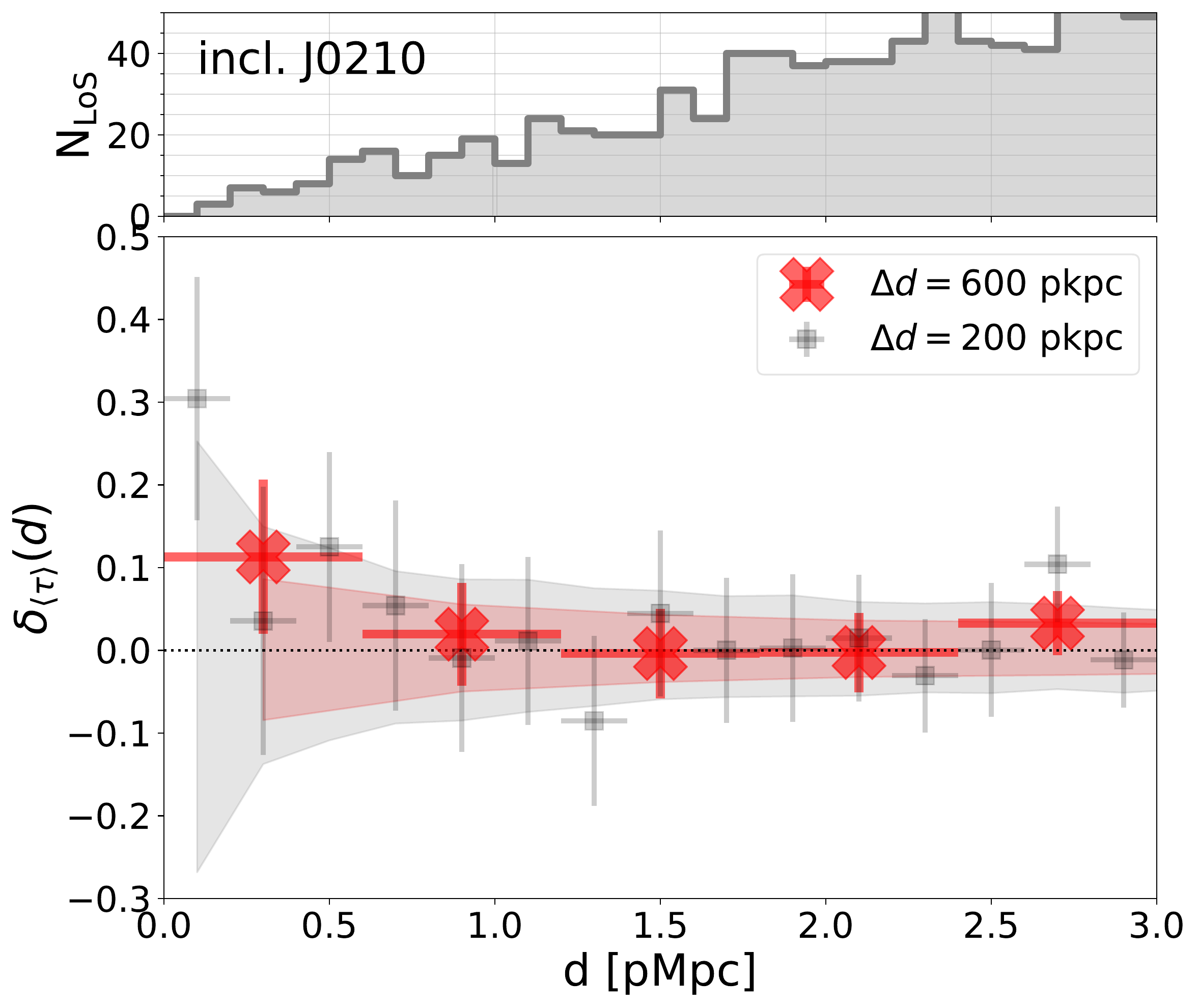}
    \caption{The fluctuation of the average \optdep~as a function of distance to LAEs, $\delta_{\left<\tau\right>}(d)$, for the case including $J0210$. 
    Errors are indicated by the 1$\sigma$ standard deviation from the 1,000 times Bootstrap resampling.
    The grey squares represent the finer bins with resolution of $200$ kpc, and the red crosses show coarse bins with resolution of $600$ kpc.
    The shaded region indicates the uncertainty for coarse bins. 
    The numbers of LoSs counted in the annulus at each step of $100$ kpc are shown in the upper panel. 
    A $30\%$ excess at a level of $2\sigma$ appears at $d<200$ pkpc indicating the detection of CGM signal around LAEs, while a tentative $13\%$ excess at $400<d<600$ pkpc shows a weaker signal in the IGM regime.}
    \label{fig:average_tau}
\end{figure}

Results derived from all the LAEs and LoSs in four fields are shown in the lower panel of Figure \ref{fig:average_tau}. 
We mainly consider the case including $J0210$ here given the same reason for CCFs, i.e., statistics is unlikely to be biased by 6 outliers out of 64 LoSs.
We check the case excluding $J0210$ in the Appendix \ref{appendix:ave_tau}, and it shows the consistent results except for a larger scatter due to the LoS number decrease.
The grey squares are for the finer bin, while the red crosses represent the coarse bin.
The error shown in y-axis is the 1$\sigma$ standard deviation from the $1,000$ times Bootstrap resampling with both the LAEs and LoSs, and the one in x-axis indicates the bin size.
The shaded regions are the 16\%--84\% ranks in the 1,000 simulations assessing random positions to the 64 LoSs with corresponding resolutions, indicating that bins outside the shaded regions are confident for inferences.
The number of LoSs counted at each finer bin can be checked with the grey step function above.

From the figure, we can learn that a $10\%$ excess beyond the error is found in the innermost region, corresponding to a scale of $<600$ pkpc.
Though the counted LoSs number in the innermost finer bin is only three, but we see a more significant $30\%$ excess at a $2\sigma$ level at $d<200$ kpc, which is the expected region distributed with CGM.
\citet{Rudie+2012} and \citet{Momose+2020b} also found the CGM signals at $d \lesssim 300$ pkpc to the star-forming galaxies that are firstly identified as LBGs.
This time, we find the indication may be also true for LAEs at a smaller distance from the statistical point of view.

In addition, the finer bins seem to also indicate a $13\%$ excess at the distance of $400<d<600$ pkpc, and given its sufficient number of LoSs, this excess is likely to be the dominant signal accounting for the $10\%$ excess in coarse bin.
However, such scale is larger than the typical region thought to be the CGM reservoir, especially with regard to LAEs which are generally less massive than LBGs.
Instead, the clustering of IGM {\sc Hi} works as a better interpretation for the excess found in $\left< \tau \right>$ at this distance.
This excess in $\left< \tau \right>$ corroborates the signal detected in CCF at $\sim 0.5$ pMpc, proving the correlation between galaxy and IGM {\sc Hi} down to around $400$ pkpc.

Between the two excess bins, $d=200-400$ kpc interestingly shows a relative valley. 
A turnover seems to appear at $d\approx400$ kpc.
The significance is very low due to the small number of LoSs counted at the bin, but we note that a similar turnover is once also reported for the $z\sim3$ LBGs by \citet{Adelberger+2003}, though theirs appears at 0.5 $h^{-1}$Mpc, or 190 kpc in our cosmology. 
The latest result based on a sample of 2,862 background galaxies has also revealed a more similar sudden dip at 70--150 pkpc \citep{Chen+2020}.
The former work has estimated that the supernova-driven outflow with the speed of $600~{\rm km}~{\rm s}^{-1}$ may cause the turnover, while the latter suggest the feature can be related to the transition phase between the inflow outside and the outflow inside, which may be related with star formation activities.
In this picture, it is possible that LAEs, the young and less massive galaxies which can be located at the shallower potential well and is active in forming stars, host the stronger outflows and cause the turnover appearing at a larger projected distance.
The current weak signal in our data still prevents us drawing any firm conclusion, but a larger sample size in the future may help to resolve this question.

We also notice there is a sudden excess of 10\% at $2.6<d<2.8$, though the coarse bin largely flatten the signal.
We do not fully understand the origin of this signal, but a non-continuous signal at such a large scale is not likely to be physically meaningful.

\section{Discussion} 
\label{sec:discuss}
Based on the results shown in Section \ref{sec:result}, we discuss their implications. 
We first make a comparison between our galaxy--IGM {\sc Hi} result with the previous literature. 
As hinted in the CCF, we find the correlation is possibly dependent on the scale.
Therefore, we further explore the scale dependence of the observed correlation.
Finally, we discuss the possible underlying physics that are related to our results on the positive correlation, correlation scale and 
the visible scatter in the \overden$-$\optdep~diagram at $z\sim2$.

\subsection{Comparison with Previous Work}
\label{sec:od2_compare}
There are already a few studies in the literature working on the correlation between galaxies and IGM {\sc Hi} on the over tens of cMpc scales at $z>2$.
The directly related work is \citet{Mukae+2017}, in which the galaxy--IGM {\sc Hi} correlation is studied by using the $Ks$-selected photo-$z$ galaxies at the redshift $2<z<3$ and the \lya~forest sample in the background quasar spectra from SDSS-III/BOSS survey.

\begin{figure*}[htb]
    \centering
    \includegraphics[width=0.8\textwidth]{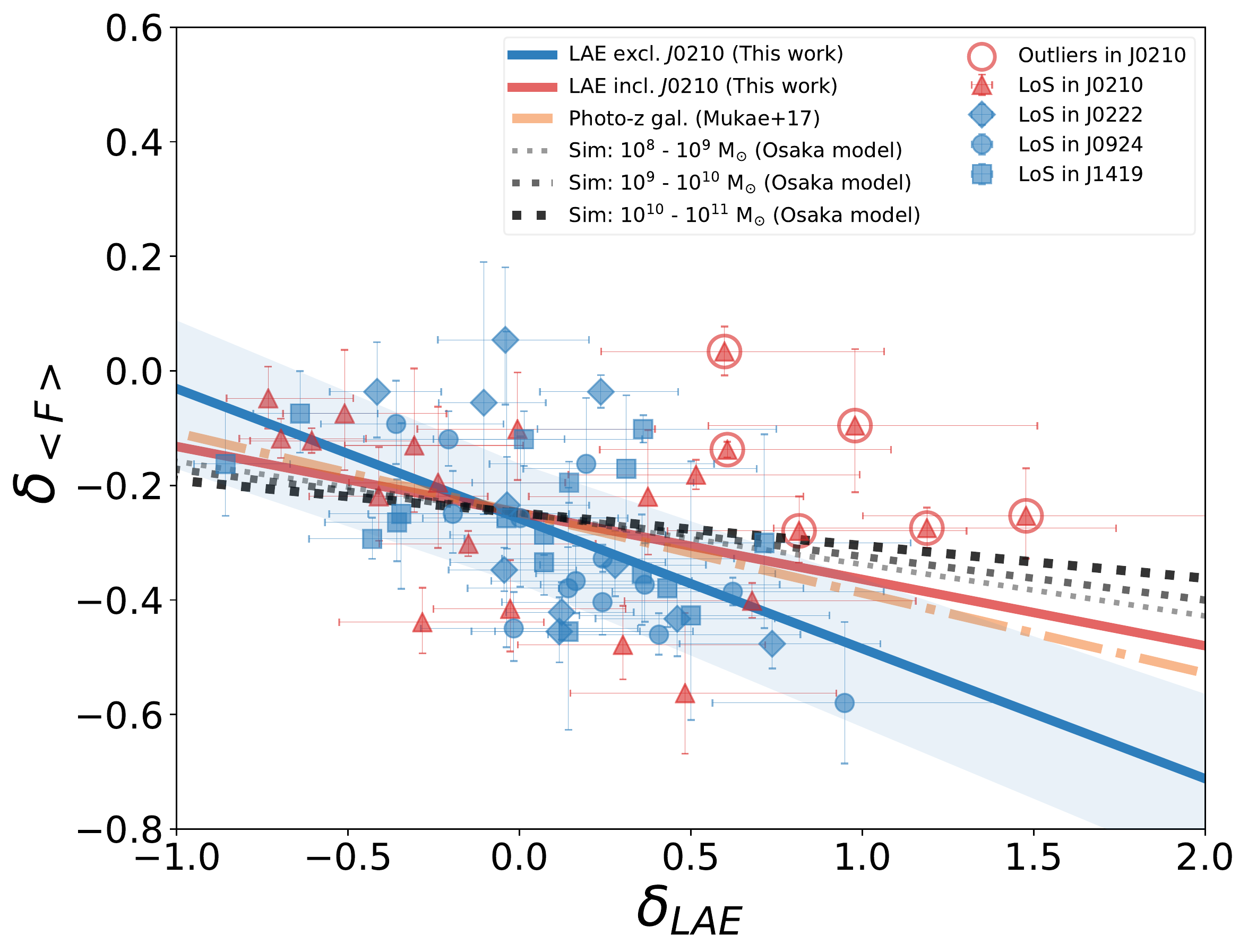}
    \caption{Transmission fluctuation $\delta_{\left< \rm F \right>}$ against the LAE overdensity \overden, similar to the Figure 2 of \citet{Mukae+2017}. 
    Symbols are same as Figure \ref{fig:od2_direct}.
    LoSs in $J0210$ are in red, while LoSs in the other three fields are painted blue.
    Outliers in $J0210$ are highlighted by red circles.
    The red (blue) solid curves are the best-fit model for the data points in(ex)cluding $J0210$.
    The orange dotted dash line is the result from \citet{Mukae+2017} using the photo-z galaxies \& 16 BOSS LoSs.
    The dotted lines are the prediction from GADGET3-Osaka model \citep{Shimizu+2019, Nagamine+2020} for galaxies with $10^8$$-$$10^9~{\rm M}_{\odot}$, $10^9$$-$$10^{10}~{\rm M}_{\odot}$ and $10^{10}$$-$$10^{11}~{\rm M}_{\odot}$.
    The shaded regions are the 16\%$-$84\% rank from the perturbation simulations for the case excluding $J0210$.}
    \label{fig:od2_fluc}
\end{figure*}

The correlation shown in their Figure 2 is physically similar to our \overden$-$\optdep~correlation, but the \lya~absorption is estimated in \lya~forest fluctuation which is defined as:
\begin{equation}
    \delta_{\left< F \right>} = \frac{\left< F\right>_{dz}}{F_{\rm cos}(z)} - 1,
    \label{eq:trans_fluc}
\end{equation}
where $\left< F\right>_{dz}$ is the transmission calculated within the redshift uncertainty $dz=0.025(1+z)$ from the spectra and the $F_{\rm cos}(z)$ is the cosmic \lya~forest mean transmission that is estimated from $F_{\rm cos}(z)=e^{-0.001845(1+z)^{3.924}}$ \citep{FG+2008}.

To compare with their results, we convert the optical depth derived in Section \ref{sec:boss_spec} into the transmission fluctuation $\delta_{\left< F \right>}$ according to the Equation \ref{eq:trans_fluc}.
The cosmic mean is also assumed to be given by the relation in \citet{FG+2008} as 0.84 at $z=2.18$.
The translated $\delta_{\left< F \right>} - \delta_{\rm LAE}$ relation from our LAEs and LoSs sample is shown in Figure \ref{fig:od2_fluc}.

The symbols of the data points are the same as Figure \ref{fig:od2_direct}, but for clarifying the different cases including or excluding $J0210$ for the fitting, we paint the LoSs in $J0210$ red and the LoSs in other fields blue.
We also make a linear fit using the Levenberg-Marquardt least-square fitting, shown as the solid lines in Figure \ref{fig:od2_fluc}, in(ex)cluding the $J0210$ correspons to the red(blue) curve.
The uncertainty of parameters is again given by the $16\%-84\%$ ranks from the 10,000 Monte Carlo simulations with perturbation.
The fitted relation for all four fields is: 
\begin{equation}
    \delta_{\left< F \right>} 
    = -0.116^{+0.018}_{-0.022} ~\delta_{\rm LAE}
      -0.248^{+0.082}_{-0.093},
    \label{eq:od2_fluc_all}
\end{equation}

Similar to Figure \ref{fig:od2_direct}, we can find the outliers in $J0210$ at the upper right in Figure \ref{fig:od2_fluc}, which is highlighted with circles.
If we exclude the LoSs in $J0210$, the relation becomes:
\begin{equation}
    \delta_{\left< F \right>} 
    = -0.227^{+0.026}_{-0.023} ~\delta_{\rm LAE}
      -0.258^{+0.096}_{-0.114},
    \label{eq:od2_fluc_e1f}
\end{equation}
which shows a steeper slope, meaning the \optdep~is more sensitive to the \overden.
We overplot the curve whose slope is $-0.14^{+0.06}_{-0.16}$ from the \citet{Mukae+2017}, with the intercept normalized at \overden$=0$.
The normalization is necessary as our tracers of absorption are not defined in the same way, which causes systematic offset reflected on the intercept.
They estimate the $\delta_{\left< F \right>}$ at the position of the highest S/N$_{\left< F \right>}$, defined as the ratio between \lya~absorption and its error, on $\sim100$cMpc scale within the redshift $2<z<3$, while we are targeting at the absorption spike based on the \optdep~on $\sim20$ cMpc scale within $2.15<z<2.20$.

Both cases in our work give the consistent slopes with the photo-z galaxies within their uncertainty, though the case excluding $J0210$ owns a larger discrepancy and is steeper.
One possible reason for the large discrepancy can be the different galaxy masses, given that photo-z galaxies are generally more massive than LAEs.
The massive galaxies are likely to form in the deeper position of the gravitational potential well, where the {\sc Hi} is abundant for building up stellar masses M$_{*}$.
In this case, the overdensity of less massive galaxies like LAEs will be systematically lower than that of the heavier populations, e.g., photo-$z$ galaxies, and thus make the $\delta_{\left< F \right>}-\delta_{\rm LAE}$ steeper.
A similar trend is also reported in a study based on the IGM tomography \citep{Momose+2020b}.
In this case, the shallower slope with $J0210$ can be explained by the possible LAE number excess in the large filament in $J0210$,
because the filament is associated with a group of quasars and this can be an indicator of the potential massive halos around the structure.
It may boost the \overden~given the same $\delta_{\left< F \right>}$, especially at the regions where the LoS outliers reside in, making the slope shallower when $J0210$ is included.

To further inspect the possibility, we refer to the results from the GADGET3-Osaka cosmological hydrodynamic simulation, which is based on the smoothed particle hydrodynamics (SPH) simulation code {\sc GADGET-3} \citep{Springel+2005} and takes full account of the star formation and supernova feedbacks \citep{Shimizu+2019}.
More details on the simulation data processing is explained in 
\citet{Momose+2020a} as well as in \citet{Nagamine+2020}, and we denote it as the {\it Osaka model} hereinafter.
The model curves for galaxies with M$_{*}$ ranging in $10^8-10^9$, $10^9-10^{10}$ and $10^{10}-10^{11}~{\rm M}_{\odot}$ with respective slopes of $-0.090 \pm 0.011$, $-0.076 \pm 0.009$ and $-0.057 \pm 0.006$ are also plotted in Figure \ref{fig:od2_fluc}, and the intercepts are again normalized at \overden$=0$, given that the absorption in the model is estimated at the fixed position, i.e., the central redshift $z=2.175$, which is different from our estimate at the absorption spike.

We do find there is an M$_{*}$-dependence of the relation slopes in the Osaka model, and the less massive galaxy population owns a steeper trend.
However, such dependence is not as sensitive as we expected and more interestingly, our fitting for the case including $J0210$ shows a good consistency with the Osaka model prediction for the galaxies with ${\rm M}_{*} \sim 10^9~{\rm M}_{\odot}$, the typical stellar masses for $z\approx2.2$ LAEs \citep{Kusakabe+2018}.
Meanwhile, the slope for the case excluding $J0210$, which was expected to be more representative of the general fields at $z\sim2$, is significantly steeper than the Osaka model.
These comparisons are likely to disprove the reason originated from the galaxy stellar masses, and the case excluding $J0210$ seems rather to be the biased case.

Another possibility can be the {\sc Hi} suppression on the \lya~emission. 
Given that our observations target at the fields with the clustering of strong IGM \lya~absorption, the \lya~emission from galaxies may get suppressed in such {\sc Hi}-rich environments before we can observe.
In this case, LAEs in the $J0210$, which is likely to contain a special structure lacking the IGM {\sc Hi} as suggested by the outlying LoSs, should be less influenced.
Meanwhile, in the other fields, the detection completeness of LAEs could be lower and the \overden~might be underestimated.
This interpretation seems to be more favored by the Osaka model prediction.
Actually, the plateau appearing in the CCF at $r \lesssim 0.6$ pMpc also supports such a possibility at least on the small scales.

However, we note that there can be some uncertainties left in the simulation models (e.g., contribution from AGNs), and our sample size is still limited for the discussions on field variation. 
In the future, we are hopefully to find out the true reason for the slope discrepancy with more HSC fields targeting at various environments. 
Follow-ups to search for H$\alpha$ emitters (HAEs) residing in the same structures, which are less biased by the radiative transfer process, can also help to robustly calibrate the $\delta_{\left< F \right>} - \delta_{\rm LAE}$ slope.

\subsection{Scale Dependence of the Correlation }
\label{sec:od2_scale}
\begin{figure*}[hbt]
    \centering
    \includegraphics[width=0.8\textwidth]{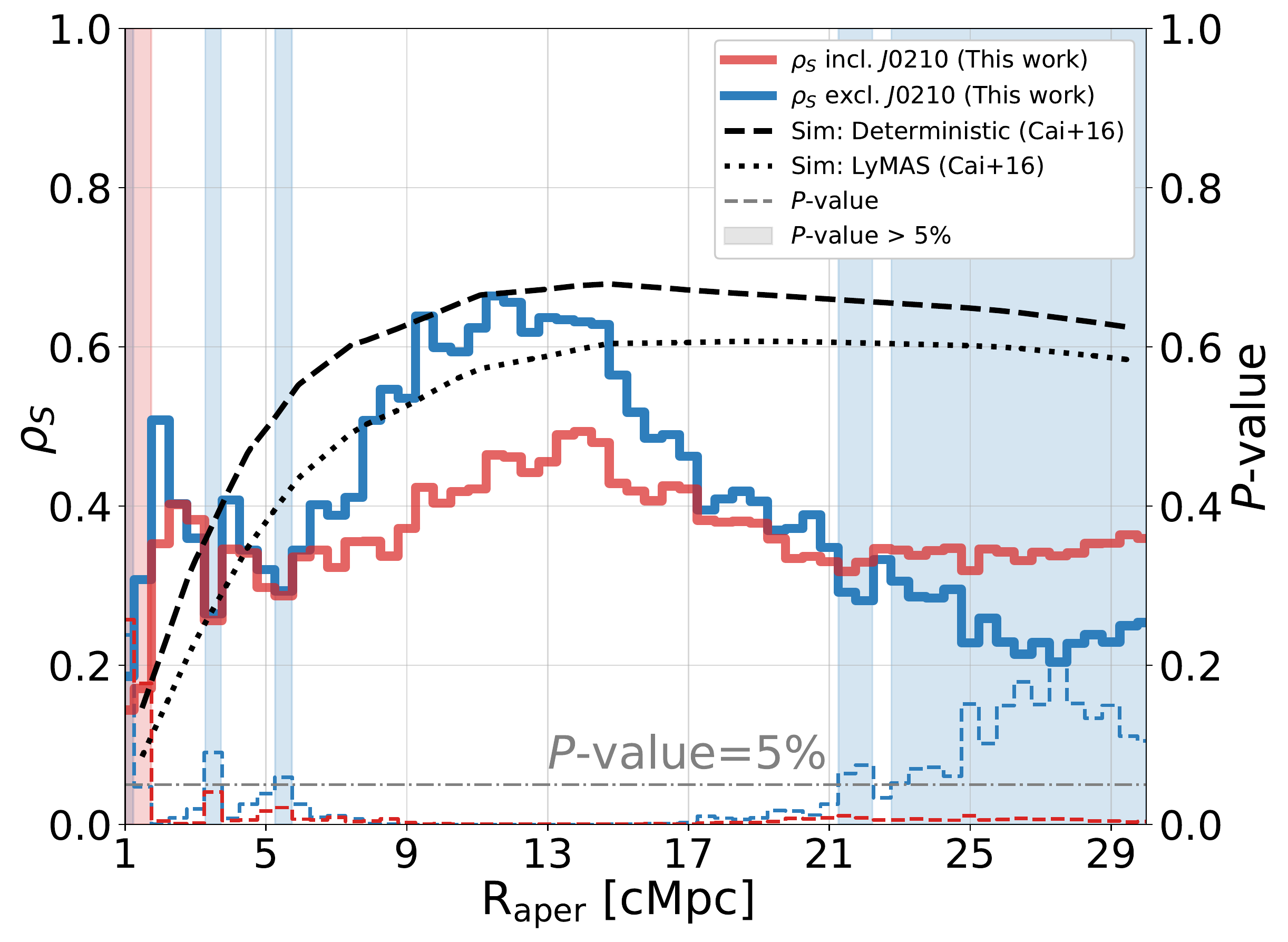}
    \caption{Scale dependence of the \overden$-$\optdep~correlation. 
    Bin size is 0.5 cMpc and the red (blue) solid curves are the Spearman's rank correlation coefficient in(ex)cluding $J0210$.
    The dash colored curves indicate the corresponding $P$-value at each bin.
    Scale range with the $P$-value $>5\%$ is masked with shaded regions, indicating that the result is not confident.
    The two models, Deterministic \& LyMAS models, from the \citet{Cai+2016} are overplotted, by scaling the boxlength in simulations with a factor or 1/2 to match the $R_{\rm aper}$.
    Both models reach the maximum at a comparable scale $R_{\rm aper} \sim 13$ cMpc.}
    \label{fig:od2_scale}
\end{figure*}
A possible scale dependence is already hinted in the CCF in Section \ref{sec:ccf_lae_los} for the LAE and IGM {\sc Hi} correlation.
To investigate the scale dependence, we perform the Spearman's rank correlation test for (\overden, \optdep) with the \overden~calculated in different aperture sizes.
We again consider the two cases, including and excluding the field $J0210$, as we already find that it may significantly alter the overdensity-based analysis in Section \ref{sec:od2_direct} and \ref{sec:od2_compare}.
The aperture size is set from 1 to 30 cMpc with a bin step as 0.5 cMpc for the radius.
We note that LoSs are kept for analysis only when $<50\%$ vicinity is masked, and this keeps a stable LoS number when the scale increases.
The result is shown in Figure \ref{fig:od2_scale}. 
The red (blue) curve shows the Spearman's rank correlation coefficients on various scales for the case including (excluding) $J0210$.
The corresponding $P$-value is shown as the dash line, and the R$_{\rm aper}$ with $P$-value $>5\%$ indicating an unconfident result is shaded.
The similar results from the deterministic as well as \lya~Mass Association Scheme (LyMAS) models in \citet{Cai+2016} are also overlaid as the black curves.
The LyMAS considers a stochastic relation described by a conditional probability distribution of the flux on the mass overdensity $\delta_{\rm m}$, based on the hydrodynamic simulations.
Note that the original box lengths in the simulations are scaled with a 1/2 to keep consistent with R$_{\rm aper}$, and this scale-match for estimating galaxy/total matter overdensity is performed on the projected plane.

It is clear that when the $J0210$ is included, the correlation keeps moderate at a level of $\rho_S \sim 0.3$ for almost all scales, while in the case excluding $J0210$, the correlation becomes strong at $r \approx 9-15$ with the $\rho_S \gtrsim 0.6$.
In both cases, we can find the scale dependence of the correlation between \overden~and \optdep\,, though the trend is much more significant when the $J0210$ is not included.
At the relatively small scale, the correlation becomes stronger with the scale increases, and it reaches a peak at $r=13\pm2$ cMpc. 
With $J0210$, the correlation shows a flatter shape when $r>13$ cMpc, while it tends to decrease at such scales if $J0210$ is rejected.
The difference again, indicates that $J0210$ may own a special structure and the existence of such structure can alter the correlation significantly in the overdensity-based analysis.
So, when doing the galaxy-IGM {\sc Hi} correlation study, a large sample size covering various types of environments should be essential.
But here, we will keep the discussions focused with $J0210$ excluded.

The CCF in Section \ref{sec:ccf_lae_los} shows that the amplitude for high \optdep~subsample keeps positive up to $4$ pMpc ($\sim12.7$ cMpc).
As the correlation shown in Figure \ref{fig:od2_scale} can be viewed as the cumulative signal within the R$_{\rm aper}$, the scale of the correlation peak agrees well with the CCF result.
Compared with the models in \citet{Cai+2016}, our result on the scale of maximum correlation is also well consistent to both of the deterministic one and the LyMAS on the projected plane, though the amplitude may be different due to the different tracers we use.
This consistency suggests the effectiveness of the current simulated cosmological models in terms of the IGM {\sc Hi} gas.

The decrease at large scale is interesting, as it is not predicted in the simulation.
Note the HSC FoV covers a region on the scale over 100 cMpc, so it is not likely to be the reason accounting for the weaker correlation on scales just over 15 cMpc. 
We also assess the mask region criterion by changing $<50\%$ to $<10\%$ when estimating the $\rho_S$. 
With this change, the analysis only uses the clean LoS sample free from the uncovered regions.
This test also shows a similar decreasing trend at $r>15$ cMpc, supporting that the decrease is unlikely to be caused by the FoV limit.

An alternative reason of the discrepancy on large scale is that the models do not only use galaxies but rather use the total matter in a defined box, which is less clustering indicated by the 
simulations that at $z\sim2$ the galaxy bias keeps decreasing towards the scale over $\sim 10~h^{-1}$cMpc \citep{Cen+2000, Springel+2018}, while the {\sc Hi} bias is almost flat at such large scale \citep{Ando+2019}.
Hence, the correlation in the models can stay strong on a larger scale, while the LAE--IGM {\sc Hi} correlation becomes weaker simultaneously.
Another possible reason suggested by the simulations in \citet{Momose+2020a} is that the signal on large scales is diluted in the projected correlation, as the uncertainty on the three dimensional separation becomes larger when R$_{\rm aper}$ increases.
In addition, more contaminants included in a larger aperture can also weaken the signal.

\subsection{Underlying Physics in the Correlation}
\label{sec:phy_cor}

\subsubsection{The Positive Correlation}
\label{sec:phy_cor_positive}
We showed that, at the redshift $z \approx 2.2$, a moderate to strong positive correlation can be found between \overden~and \optdep~on a scale of $r=10$ cMpc. 
Such a correlation suggests that galaxies are clustering in a region associated with large amount of {\sc Hi} gas.
This correlation is found to be scale dependent, and the peak locates at R$_{\rm aper}\sim 13$ cMpc. 
The correlation seems to be natural in a simple picture that IGM {\sc Hi} gas tends to be accumulated in the deeper potential wells which are inhabited by the more massive halos.
The condensed {\sc Hi} gas then triggers star formation, and stars and galaxies will emerge at the same region.
Especially at $z\sim2$, such activity is extremely intensive according to \citet{Madau+2014}. 

However, when detailed processes are taken into account, the situation becomes complicated.
Hot massive stars can emit ionizing photons with energy $>13.6$ eV, and once they succeed to escape from the host galaxies, the surrounding  {\sc Hi} gas in the surrounding IGM will be ionized.
Such process can be more active for the case of LAEs used in our work, which are thought to be a population of young star-forming galaxies. 
Our result indirectly suggests that the escape fraction of ionizing photons from LAEs at $z\sim2$ or their SFR is still not high enough to fully ionize the IGM {\sc Hi} gas on the scale of several cMpc.
Feedback from supernovae or the potentially inhabiting AGNs can also be possible to blow off the surrounding gas to more distant regions, though how powerful such processes can be is still under debate and it is not clear up to which scale they can affect.

Some literature has explored the two point CCFs between \lya~absorbers and galaxies in the lower-$z$ universe \citep{RW+2006, Chen+2009, Tejos+2014}, in which the correlations are confirmed under the redshift $z\lesssim1$. 
But at $z\gtrsim2$, such correlation can be only constrained with limitations in either bright galaxy population, small survey area or small LoS sample size in a limited number of works \citep{Adelberger+2003, Rudie+2012, Mukae+2017}.
Our result confirms the correlation between IGM {\sc Hi} and galaxies with rest-frame UV magnitude down to M$_{\rm UV} \approx -18$ estimated from observed $g$-band, even at the redshift $z\approx2.2$ where the star formation and feedback processes can be very active.
The result shows a rough consistency with \citet{Mukae+2017} based on photo-$z$ galaxies, but a factor of $\sim4$ larger sample size in both of the LoS number and survey area makes the statistics more robust with various overdense environments.

The identified positive correlation is found up to 4 pMpc (or 13 cMpc) from the CCF analysis (or varying the aperture size for \overden~in the \overden--$\delta_{\left< F \right>}$ correlation), and down to at least 400 pkpc (or 1.3 cMpc) from the average \optdep~profile centered at LAEs.
This suggests the ionization or feedback from galaxies (LAEs) is not sufficient enough to cancel out the gravitational effects on large scale.
This indicates that IGM {\sc Hi} still traces well LSS at $z\sim2$ on the scale $1.3 \sim 13$ cMpc, though with large scatter.
Alternatively, the correlation can also be a result of additional inflow providing exceeding pristine {\sc Hi} gas \citep{DB+2006, Tumlinson+2017}.
\citet{Turner+2017} suggests the observed redshift–space distortions in the KBSS survey \citep{Rakic+2012} are predominantly caused by infall, which proves gas inflow can alter observables up to a scale of 5 pMpc.
The two possible scenarios can either or both reproduce our results and cannot be distinguished at this point. 
But it will be possible to answer this question by comparing our results with numerical simulations in the future.

Also, we still have little knowledge on how well the LAEs trace the underlying structures, especially in our fields which are expected to be associated with neutral IGM gas.
Physical similarity between LAEs and non-LAEs at $z\sim2.2$ is hinted in \citet{Hathi+2016}, and \citet{Shimakawa+2017} also find the overdense regions traced by LAEs and HAEs show good consistency on the scale of $>1$ cMpc, indicating that LAEs can be a good structure tracer on large scale.
However, as reported in \citet{Shi+2019}, LAEs and LBGs do possibly trace different structures formed in different period or in different dynamic status.
Especially on small scale of $\lesssim300$ pkpc, or $<1$ cMpc, tentative deficit is always found for LAEs, both in this work hinted by the plateau shape in the CCF and in the literature, e.g., LAE number deficit in a protocluster core \citep{Shimakawa+2017} or at the center of the massive overdensity \citep{Cai+2017a}, and the possible \lya~suppression in galaxy overdense regions \citep{Toshikawa+2016}.
This indicates that the LAE may be not a good tracer of the highest overdensity regions.

But for a statistical study on large scale, LAEs still work as the best tool with a well constrained redshift $\Delta z \approx 0.04$ and Subaru/HSC can map the objects with high efficiency.
In the future, we will perform the NB imaging with NIR instruments like Subaru/MOIRCS, on which the appropriate NB2083 ($\lambda_0 = 2.083~\mu{\rm m}$) filter is installed, to select the resonance-free HAEs to figure out the performance of the LAE tracers.

\subsubsection{The Scale Dependence of Correlation}
\label{sec:phy_cor_scale}
Our results on the scale dependence of the correlation was discussed in Section~\ref{sec:phy_cor_positive}. 
The Subaru/HSC allows us to map extended structures as well as their environments up to a scale over 100 cMpc, down to a depth $L_{\rm Ly\alpha}\approx 2 \times 10^{42}~{\rm erg}~{\rm s}^{-1}$ at $z\approx2.2$.
It makes our study unique for robustly confirming correlation on a large scale of several tens of Mpc in comoving at $z>2$.

As mentioned previously, there are already some studies working on the CCF between \lya~absorbers and galaxies at $z<1$ \citep{RW+2006, Chen+2009, Tejos+2014}, and the CCFs provide us the information for both correlations and their effective scales.
Given that the galaxy populations and the \lya~absorption systems used among our works are not identical, it is hard to directly compare the CCF amplitude and the resulting clustering length $r_0$.
Nevertheless, the $r_{\rm up}$, defined here as the upper limit of the scale to identify the positive signal, can be still instructive.
From the CCFs, we find an underlying redshift evolution of the correlation scale by combining \citet{RW+2006}, \citet{Tejos+2014}, whose CCFs also extend over 10 cMpc, with our result.
We find that: (1) at $z\lesssim 0.04$, the CCF between \lya~absorbers with {\sc Hi} column density ranging in $10^{12.5} \lesssim N_{\text{\sc Hi}} \lesssim 10^{15} {\rm cm}^{-2}$ and HIPSS galaxies shows a strong positive signal up to 10 $h^{-1}$cMpc \citep{RW+2006}, i.e., $r_{\rm up} \sim 15$ cMpc, slightly larger than our upper limit $r_{\rm up} = 13\pm3$ cMpc;
(2) while at $0 \lesssim z \lesssim 1$, the signal of CCF between \lya~absorption systems with {\sc Hi} column density ranging in $10^{14} \lesssim N_{\text{\sc Hi}} \lesssim 10^{17} {\rm cm}^{-2}$ and galaxies, can be only found up to $r_{\rm up} \sim 7$ cMpc \citep{Tejos+2014}, significantly smaller than ours.

The $r_{\rm up}$ decreases from $z>2$ to $0 \lesssim z \lesssim 1$, and then increases again towards $z=0$. 
This interestingly shows a consistency with the varying trend of the correlation length of galaxy clustering \citep{Baugh+1999, Springel+2018}.
It supports a physical picture that the redshift evolution of galaxy-IGM {\sc Hi} correlation may follow a similar pattern of the galaxy clustering.

\subsubsection{The Scatter of Correlation}
\label{subsec:phy_cor_scatter}
The scatter can also be an important indicator of the underlying structures.
As the Figure~\ref{fig:od2_direct} and \ref{fig:od2_fluc} show, data points are distributed with a large scatter.
It may originate from the uncertainties in our measurements.
We summarize the possible factors here. 
First, regarding the overdensity measured in our work, we can only map the LAEs on projected plane while an uncertainty of $\sim 60$ cMpc is left along the redshift direction, and the aforementioned scales are all in transverse separation instead of in comoving volume.
Additionally, we are not sure how much bias LAEs are introducing, as we already discussed in Section \ref{sec:phy_cor_positive}.
As $J0210$ changes the statistical results very much, structures with field-to-field variation may exist.
Regarding the LoSs sample, though the CoSLAs have been carefully checked to exclude DLAs or LLSs, low \optdep~LoSs can be still possibly contaminated by these systems. 
But even if we only focus on CoSLAs (see LoSs with \optdep~$\gtrsim 0.6$ in Figure \ref{fig:od2_direct}), we can still find a large scatter, just like what \citet{Miller+2019} report in their simulations with both the high spatial and mass resolutions.
This indicates there should be some intrinsic origins.

The scatter can be coincidences that happen when LoSs pass through a gas filament, a large void or an orthogonal filament with low density. 
According to the simulation in \citet{Mukae+2017}, which also find a large scatter on their correlation, it may indicate the outliers in $J0210$ penetrate a galaxy overdensity associated with a gas filament lying on a transverse direction to the LoS by chance.
In addition to the morphological origin, the radiation from galaxies may preheat the diffuse IGM {\sc Hi} in the most overdense region, causing the scatter.
This scheme is suggested by \citet{Mawatari+2017}, where \lya~absorption is found to be associated with a $z\approx 3.1$ overdensity SSA22 on a scale $\sim50$ cMpc overall, but not dependent on local overdensity.

Actually, three outliers in $J0210$ located at regions with \overden~$\gtrsim1.0$ in the $r_{\rm aper}=10$ cMpc aperture and \optdep~$\lesssim0.4$ are just likely to reside in the regions that are abundant with galaxies but in deficit of cold {\sc Hi} gas, similarly to the environments mentioned above. 
A special system found in the IGM tomography also shows the similar characteristics \citep{Lee+2016}.
By further studying such cases in the scatter, we may be able to find more ideal laboratories for testing the theories of galaxy evolution and their interplay with IGM {\sc Hi} in the extreme environments at $z\approx2$.

\section{Summary} 
\label{sec:summary}

In this paper, we perform deep NB387 and $g$-band imaging with the 8.2-m Subaru/HSC on the fields following the similar technique used in MAMMOTH project, which are preferentially traced by the group of strong Ly$\alpha$ absorbers selected from the full (e)BOSS database. 
Using the narrow-band images, we select out LAE candidates at $z=2.18$ and construct the \overden~maps. 
To estimate the IGM {\sc Hi} overdensity, we use the (e)BOSS LoS data to calculate the \optdep~at the same redshift.
Based on the \overden~and \optdep~data, we perform correlation analyses to study the galaxy$-$IGM {\sc Hi} correlation up to a scale of $\sim100$ cMpc. 
In addition, we also examine the correlation on CGM scales down to 200 pkpc based on the statistical sample.

The results achieved are summarized as follows:
\begin{enumerate}[leftmargin=0.5cm]

    \item We construct the LAE overdensity maps for four HSC fields traced by IGM {\sc Hi} at $z=2.18$, with a total of 2,642 LAE candidates detected down to $L_{\text{Ly}\alpha} \approx 2 \times 10^{42}~{\rm erg~s}^{-1}$ over a survey area of $5.39$ deg$^2$. 
    The selected LAE candidates reside in a variety of environments, including the filaments, sheets and clumps.
    The $J0210$ field, which is associated with 11 quasars within $\sim40 \times 40$ cMpc$^2$ and $\Delta z\approx0.05$, is found to be associated with a large LAE filament extending for about 100 cMpc, one of whose nodes reaches the overdensity significance of $>6\sigma$.
    
    \item We find a moderate to strong correlation between the \overden~and \optdep~based on 64 LoSs from SDSS/(e)BOSS, which shows a rough consistency with the results in \citet{Mukae+2017}, though the $\delta_{\left< F \right>}-\delta_{\rm LAE}$ slope is steeper when we exclude the field $J0210$. 
    Based on the comparison with the Osaka simulation model \citep{Shimizu+2019, Nagamine+2020}, the discrepancy is unlikely to be caused by different stellar masses, but rather due to the suppression of \lya~emission in high {\sc Hi} density regions.
    We further find that the correlation depends on the scale of \overden~estimate.
    The peak of the correlation is located around R$_{\rm aper}=13\pm2$\,cMpc.
    
    \item By dividing the LoSs into high and low \optdep~subsamples with a criterion of \optdep~$=0.5$, the cross-correlation analysis shows a significant correlation signal up to $4\pm1$\,pMpc ($\sim 12.7\pm3.2$\,cMpc).
    The result clearly suggest that LAEs tend to reside in the gas-rich regions, which is indicated by the high \optdep~in the background LoS, and avoid the low \optdep~area where the {\sc Hi} is deficient.
    The plateau shape at $r\lesssim600$\,pkpc suggests the offset of LAEs and IGM {\sc Hi} on the small scale.
    
    \item The analysis of the average \optdep~profile centered at LAEs can trace the absorption signal down to a scale of 200\,pkpc.
    We find a 30\% excess at $d<200$\,kpc, though only with three LoSs counted, indicating the statistical detection of the CGM signal around LAEs.
    We also detect a signal of $13\%$ excess at $400-600$\,pkpc
    that is supposed to be in the IGM regime, supporting the IGM signal detection down to $\sim400$ pkpc.

    \item The positive correlation indicates that, at $z\sim2$, neither ionization nor supernova/inhabiting AGN feedback from LAEs are sufficient to erase the gravitational effects on galaxy$-$IGM {\sc Hi} correlation, or alternatively, the exceeding inflows keep supplying {\sc Hi} gas from a very large scale to the surrounding environment of galaxies.
    
    \item By comparing our correlation scale with CCFs between \lya~absorbers and galaxies at $z < 1$ \citep{RW+2006, Tejos+2014}, we find that the redshift evolution of galaxy$-$IGM {\sc Hi} correlation may follow the evolution
    of galaxy clustering.
    
    \item We also find a large scatter in the \overden~$-$ \optdep~correlation. 
    Referring to the simulation in \citet{Mukae+2017}, outliers may be the cases that LoSs penetrate regions with specific morphological arrangement.  
    In the high overdensity end, exceeding ionization and pre-heating process may be the reasons for the deficit of cold IGM {\sc Hi}, just like the $z=3.1$ protocluster in SSA22 field \citep{Mawatari+2017}.
    
\end{enumerate}

The project is still on-going for obtaining more LAEs in different fields and more LoSs in the overdense regions to strengthen the statistical robustness, so that we can compare the observables with simulations to tell the models of structure formation and evolution in terms of IGM {\sc Hi} in the future.
The upcoming Subaru/Prime Focus Spectroscopy (PFS) will be of high efficiency to make the spectroscopic confirmation for our LAE candidates, and also will provide us a good chance to perform IGM tomography in various environments, especially those with coherently distributed IGM {\sc Hi} and overdensities.


\acknowledgments
\section*{Acknowledgments}
We thank the anonymous referee for the kind comments and suggestions, which have helped to improve this paper a lot.
We also thank Drs. Ishikawa, S., Iwata, I., Kusakabe, H., Kakuma, R., Momose R., Nakanishi K., Onodera, M. and Ouchi, M. 
for their discussions and instructive advice on our works.

This work was supported in part by the Graduate University for Advanced Studies, SOKENDAI.
This work is based on data collected at Subaru Telescope, which is operated by the National Astronomical Observatory of Japan. 
This paper makes use of software developed for the Large Synoptic Survey Telescope. We thank the LSST Project for making their code available as free software at \url{http://dm.lsst.org}.
Numerical simulations were carried out on the Cray XC50 at the Center for Computational Astrophysics, National Astronomical Observatory of Japan, and the {\small OCTOPUS} at the Cybermedia Center, Osaka University as part of the HPCI system Research Project (hp180063, hp190050). 
This work is supported in part by the JSPS KAKENHI Grant Number JP17H01111, 19H05810 (K.N.).
KN acknowledges the travel support from the Kavli IPMU, World Premier Research Center Initiative (WPI), where part of this work was conducted.

The authors wish to recognize and acknowledge the
very significant cultural role and reverence that the summit of Maunakea has always had within the indigenous
Hawaiian community. We are most fortunate to have the
opportunity to conduct observations from this mountain.

{\it Facility:} Subaru (HSC)

\appendix

\section{Correction of photometric zero-point}
\label{appendix:zp_correct}

We notice that there is a systematic offset in the Equation \ref{eq:photo_check} for NB387, and we should introduce a constant $C_{\rm metal}$ for correction.
Because the colors between NB387, $g$ and $r$ are influenced by the $4,000$ \AA~$break$, which is sensitive to the metallicity \citep{Kauffmann+2003}. 
The Pickles templates are mainly constructed from the stars with solar metallicity \citep{Pickles+1998}, while the number of star references used in the {\it hscPipe} has the peak around $19<g<21$, and so, tend to be the metal-poor halo stars that are more distant to us at high Galactic latitude.
This difference may cause a systematic bias.

\begin{figure}[hth!]
    \centering
    \includegraphics[width=0.45\textwidth]{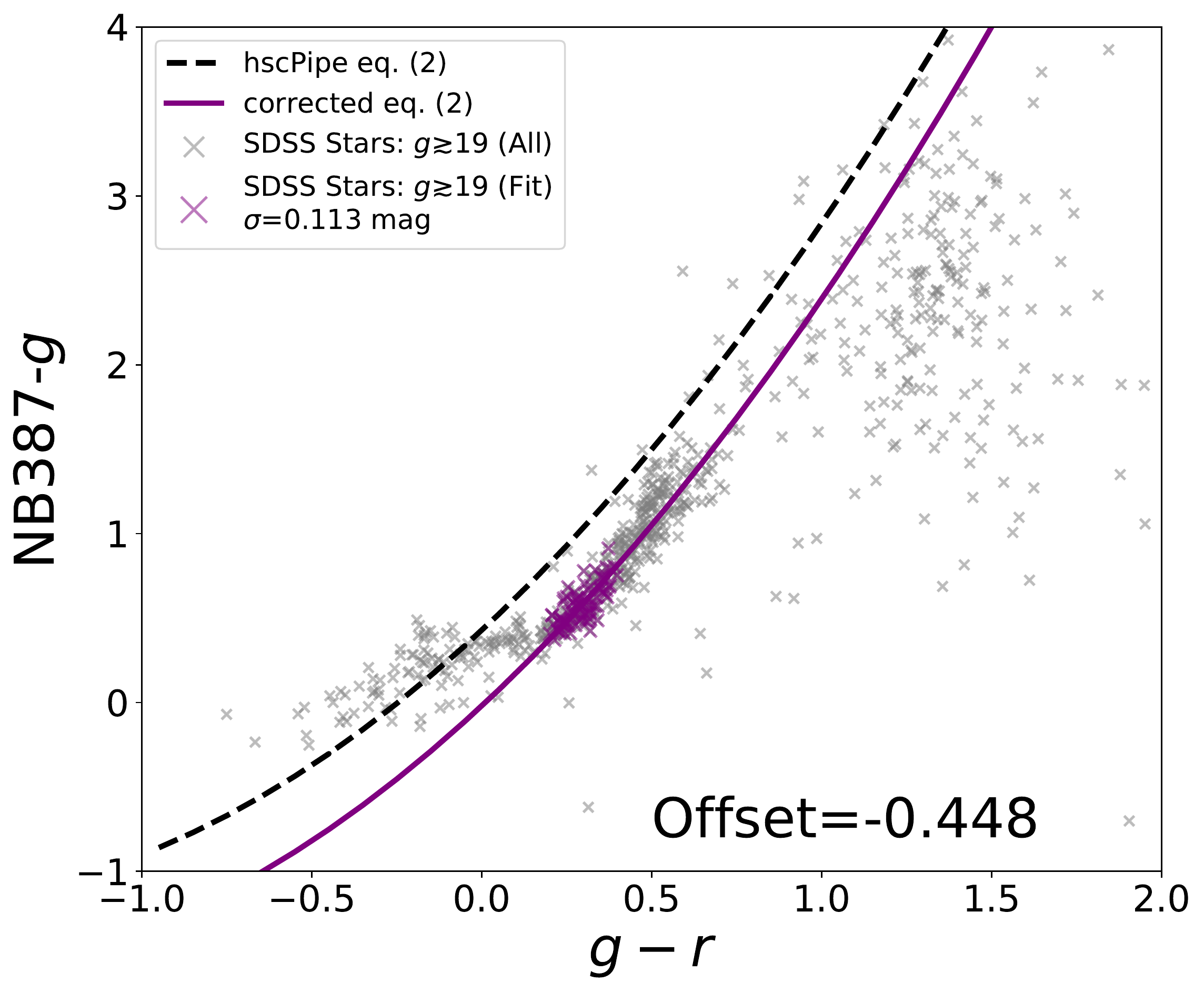}
    \includegraphics[width=0.45\textwidth]{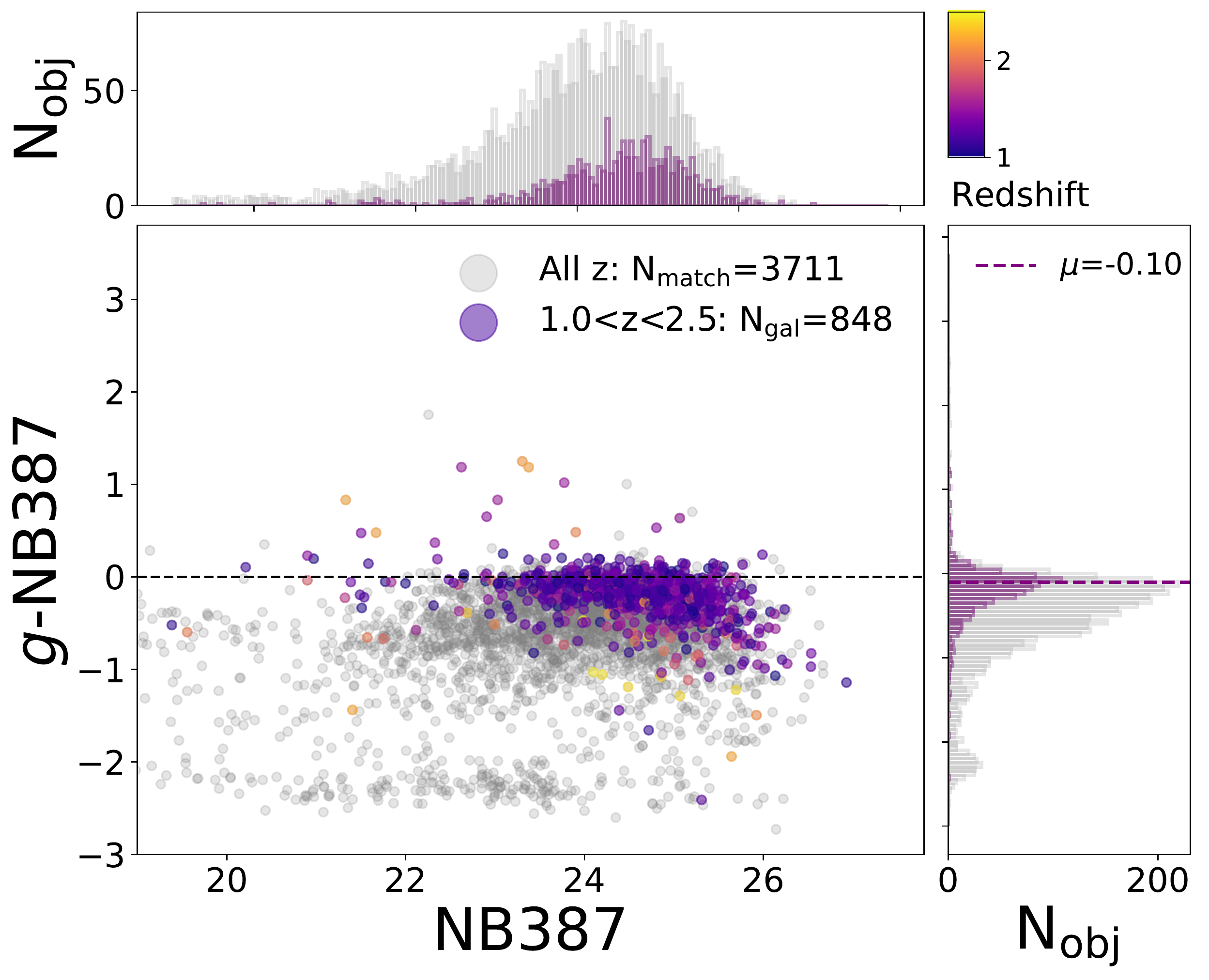}
    \caption{({\it Left}) The predicted ${\rm NB387 - g}~vs.~g-r$ diagram for homogeneously selected SDSS stars with $g\gtrsim19$.
    The grey crosses are all the selected stars, and the purple ones are those with $0.2<g-r<0.4$ after visual inspection, which are used for fitting the correction factor $C_{\rm metal}$.
    The black dash curve is the Equation \ref{eq:photo_check} from {\it hscPipe}, and the purple solid curve is the corrected relation with $C_{\rm metal} = -0.448$.
    ({\it Right}) The $g-{\rm NB387}~vs.~{\rm NB387}$ diagram for the 2\arcsec~cross-matches between the CHORUS objects (Inoue et al. 2020, submitted) and the DEIMOS 10K catalog \citep{Hasinger+2018}.
    The grey dots are all the 3,711 matches with flag $q>1$, suggesting a robust spectral redshift $z_{\rm spec}$ measurement, and the dots coded with the hot map are the 848 high-$z$ matches with $1.0<z_{\rm spec}<2.5$, and the hotter means the higher redshift.
    }
    \label{fig:zp_corr_sdss}
\end{figure}

To estimate the $C_{\rm metal}$, we homogeneously select the faint stars ($g\gtrsim19$) with S/N $>3$ spectra at the NB387 wavelength range from the SDSS database around the COSMOS field, whose Galactic latitude is comparable to our case.
Then we calculate the predicted HSC/NB387, PS1/$g$-band and PS1/$r$-band magnitudes for these stars by taking their total transmission curves into accounts.
These stars are plotted as the grey crosses in Figure \ref{fig:zp_corr_sdss}.
To keep consistency with the fitting in {\it hscPipe} and also to reduce the fitting uncertainty, we only use stars with $0.2 < g-r < 0.4$, which shows the smallest scatter in the relation.
Most of the selected stars are flagged as the SEGUE targets in the SDSS \citep{Yanny+2009}.
For robust estimate, we perform visual inspection on each spectra of all these stars to discard those with weird features at the NB387 wavelength range.
After this check, stars used for the zero-point correction is plotted as the purple crosses in Figure \ref{fig:zp_corr_sdss}.
We use these realistic stars, instead of the Pickles templates, to fit the relation shown in Equation \ref{eq:photo_check} and the $C_{\rm metal}$ is estimated as -0.448.
The original relation fit from {\it hscPipe} is shown as the black dash curve, and the corrected one is shown as the purple solid curve. 

When fitting the Equation \ref{eq:photo_check}, the scatter of references is large in the case of NB387, making the fitting uncertainty as large as 0.2 mag and thus causing a field-to-field variation.
We do the more subtle calibration for it by introducing another constant $C_{\rm fit}$.
We first select out the extended sources with $23.5 < {\rm NB387} < 24.5$, which are most likely the high-$z$ galaxies that are free from the $4, 000$ \AA~$break$ in $g$-band, in each field.
Then the field dependent $C_{\rm fit}$ is estimated by adjusting the $g - {\rm NB387}$ of these sources to -0.10, the expected mean color of $1 < z < 3$ galaxies given their typical UV slope \citep{Kurczynski+2014}. 

The $g - {\rm NB387}=-0.10$ can also be verified by utilizing the HSC/NB387 data from CHORUS survey (Inoue et al. 2020, submitted) and the spectral redshift $z_{\rm spec}$ from DEIMOS 10K spectroscopic survey catalog \citep{Hasinger+2018} in the COSMOS field.
We first cross-match the CHORUS objects with the spectroscopic catalog within a 2\arcsec~aperture, and there are 3,711 matches with flag $q>1$ suggesting the good spectroscopic redshift measurement.
The $g-{\rm NB387}~vs.~{\rm NB387}$ with $C_{\rm metal}$ correction of these objects are plotted as the grey dots in the right panel of Figure \ref{fig:zp_corr_sdss}.
We pick out all the 848 high-$z$ galaxies with $1.0<z<2.5$ from the matched catalog, which are coded by the heat map in the figure, to measure the mean of the $g-{\rm NB387}$ in a dual-Gaussian distribution, as the faint objects are likely in a flatten distribution due to photometric errors.
The result for the main sequence peak is $\mu=-0.10$, being consistent with the expected color of high-$z$ galaxies. 
This consistency also validates the $C_{\rm metal}$ as the confident correction, and because the CHORUS NB387 data is observed in excellent conditions and has a plausible depth, it is reasonable to use the suggested value $-0.10$ for calibrating the $C_{\rm fit}$ in each of our fields in this paper.
The resulting $C_{\rm fit}$ fluctuates in $-0.002 - 0.191$ mag among the four fields, which is consistent within the fitting uncertainty of $0.2$ mag.

\section{Test of the CCF results}
\label{appendix:ccf_subsample}

\begin{figure}[hth!]
    \centering
    \includegraphics[width=0.8\textwidth]{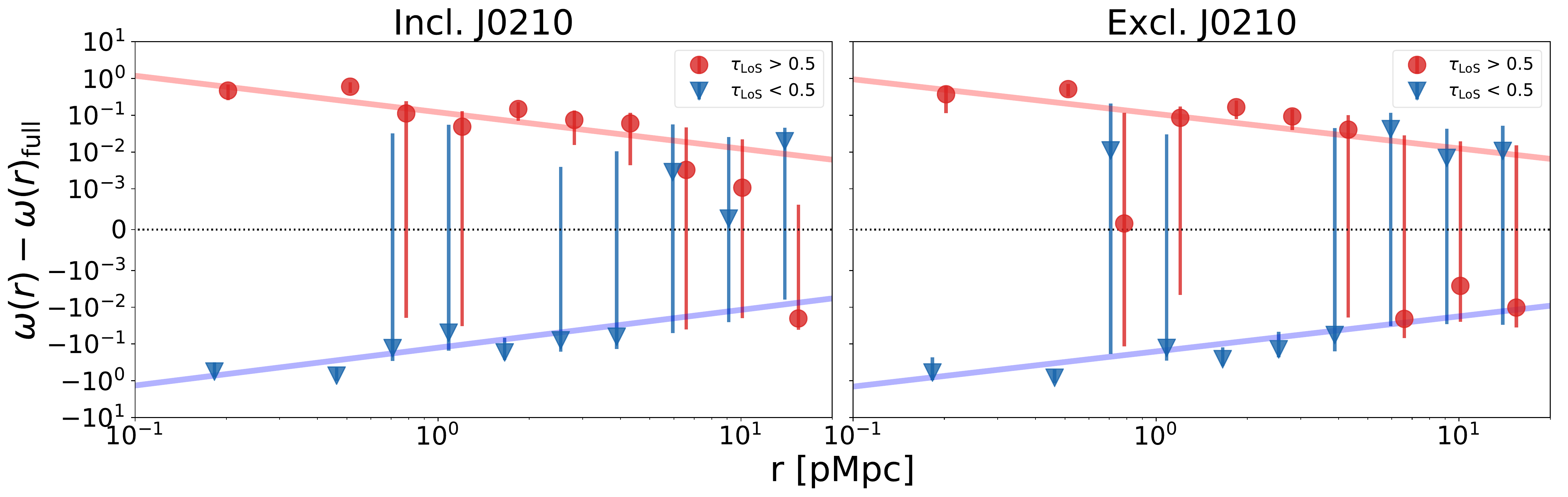}
    \caption{Relative CCFs between LAEs and LoSs for the high \optdep/low \optdep~subsamples in log scale for checking results for the cases including ({\it left panel}) and excluding ({\it right panel}) $J0210$, similar to the inset figures in the right panel of Figure \ref{fig:ccf_linear_log}. 
    Bins are set in log scale with right boundary from 0.4 pMpc to 18.3 pMpc. 
    Red points and curves are the \optdep~$>0.5$ subsample and corresponding fit power law model, while blue points and curves are the \optdep~$<0.5$ subsample. 
    The fit parameters can be checked in Table \ref{tab:ccf_param}, and they are not significantly changed between the two cases.}
    \label{fig:ccf_origin}
\end{figure}

The cross-correlation function (CCF) presented in Section~\ref{sec:ccf_lae_los} may show some variation by changing the sample size.
Here, we first test the difference between cases including and excluding the field $J0210$.
The results in log scale are shown in Figure \ref{fig:ccf_origin}, where the left panel shows results including $J0210$ while the right one excludes it.
From the comparison, no significant change in the results is found when we exclude $J0210$, except for one bin around 0.8\,pMpc and generally larger errors, probably due to a smaller sample size.
There is also not much variation in the clustering strength indicated by the $r_0$, which is summarize in Table \ref{tab:ccf_param}.

\begin{figure}[htb!]
    \centering
    \includegraphics[width=0.8\textwidth]{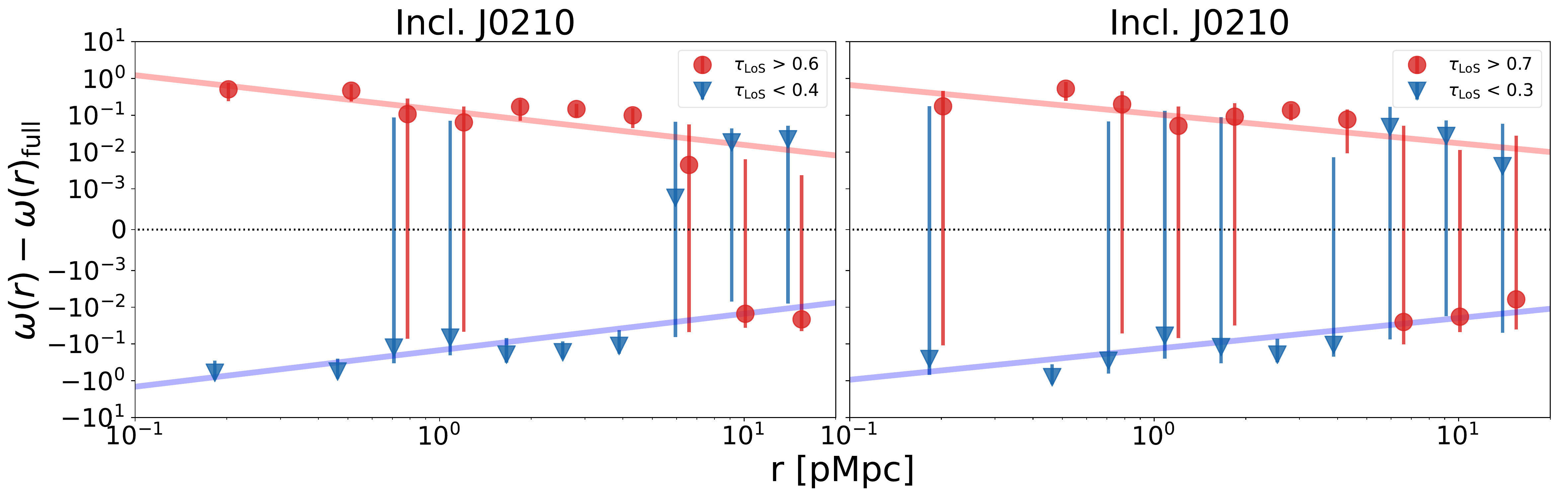}
    \caption{Relative CCFs between LAEs and LoSs for the high \optdep/low \optdep~subsamples in log scale for checking results varying the subsample criteria.
    Symbols are similar to the Figure \ref{fig:ccf_origin}. 
    Both figures are the results for the case including $J0210$.
    {\it Left panel}: for the subsamples \optdep~$>0.6$/\optdep~$<0.4$; {\it right panel}: for the subsamples; \optdep~$>0.7$/\optdep~$<0.3$. 
    The fit parameters can be checked in Table \ref{tab:ccf_param}, and they do not significantly changed, either.}
    \label{fig:ccf_subsample}
\end{figure}

\begin{table*}[htbt!]
\centering
\begin{tabular}{@{}c|cccc|cccc@{}}
\toprule
Fields & \optdep~$>$ & N$_{\rm LoS}$ &$\gamma$ & r$_0$ & \optdep~$<$ & N$_{\rm LoS}$ & $\gamma$ & r$_0$ \\ 
$[1]$ & $[2]$ & $[3]$ & $[4]$ & $[5]$ & $[6]$ & $[7]$ & $[8]$ & $[9]$ \\ \midrule
Incl. $J0210$ & 0.5 & 30 & $0.99^{+0.54}_{-0.17}$ & $0.12^{+0.05}_{-0.03}$ & 0.5 & 34 & $1.03^{+0.83}_{-0.21}$ & $0.13^{+0.06}_{-0.02}$  \\
Excl. $J0210$ & 0.5 & 23 & $0.94^{+0.66}_{-0.16}$ & $0.09^{+0.05}_{-0.03}$  & 0.5 & 19 & $0.96^{+0.95}_{-0.19}$ & $0.15^{+0.10}_{-0.05}$ \\ 
Incl. $J0210$ & 0.6 & 21 & $0.95^{+0.58}_{-0.13}$ & $0.12^{+0.07}_{-0.03}$ & 0.4 & 24 & $0.99^{+0.92}_{-0.16}$ & $0.15^{+0.09}_{-0.04}$ \\ 
Incl. $J0210$ & 0.7 & 13 & $0.79^{+1.97}_{-0.11}$ & $0.06^{+0.07}_{-0.06}$ & 0.3 & 13 & $0.84^{+2.07}_{-0.15}$ & $0.09^{+0.14}_{-0.09}$ \\
\bottomrule
\end{tabular}
\caption{The parameters of CCF power law fitting for different subsamples. 
$[1]$: cases regarding field $J0210$; 
$[2]$: high \optdep~criterion;
$[3]$: number of LoSs in the high \optdep~subsample;
$[4]$: $\gamma$ fit for high \optdep~subsample;
$[5]$: $r_0$ fit for high \optdep~subsample;
$[6]$: low \optdep~criterion;
$[7]$: number of LoSs in the low \optdep~subsample;
$[8]$: $\gamma$ fit for low \optdep~subsample;
$[9]$: $r_0$ fit for low \optdep~subsample;}
\label{tab:ccf_param}
\end{table*}

Because the definition of subsamples is based on a criterion, i.e., \optdep~over/lower than 0.5, which is kind of arbitrary, we also test whether varying the criteria will change the result or not.
We divide LoSs into other subsamples with \optdep~$>0.6$/\optdep~$<0.4$, \optdep~$>0.7$/\optdep~$<0.3$ respectively, to ensure sample size for each subsample is comparable as of 21/24 and 13/13.
The results are shown in the Figure \ref{fig:ccf_subsample}.
When we compare the results with the one shown in the left panel of Figure \ref{fig:ccf_origin}, no significant changes can be found in the trend of CCFs, except for a larger uncertainty because of smaller sample size. 
The fitted $r_0$ is summarized in Table \ref{tab:ccf_param}, and they are still of the same order of $\sim 0.1$\,pMpc scale.
These consistency prevents the galaxy$-$IGM {\sc Hi} correlation up to a scale of $\sim4$ pMpc at $z=2.2$ hinted in the CCF analysis from a coincidently defined criterion.

\singlespace

\section{Average Optical Depth Profile excluding $J0210$}
\label{appendix:ave_tau}

In Section \ref{sec:average_tau}, given the importance of the LoS number for the statistics when inspecting small scales, we mainly discuss the case with $J0210$, which contains a large filament with a group of quasars associated.
Here, we show the result for the case excluding $J0210$, and we do not find a significant change on the general $\left< \tau \right>$ varying trend along the distance to LAEs at the inner region that is discussed in Section \ref{sec:average_tau}, although the scatter is larger due to a smaller number of LoSs.
This supports our assumption that the statistics, such as the $\left< \tau \right>$, is unlikely to be affected by the six outliers out of 64 LoSs.
We note that two finer bins at $\sim2.7$ pMpc show the tentative excess, more significant than the case including $J0210$, although the coarse bin still shows a weak signal.

\begin{figure}[htb!]
    \centering
    \includegraphics[width=0.5\linewidth]{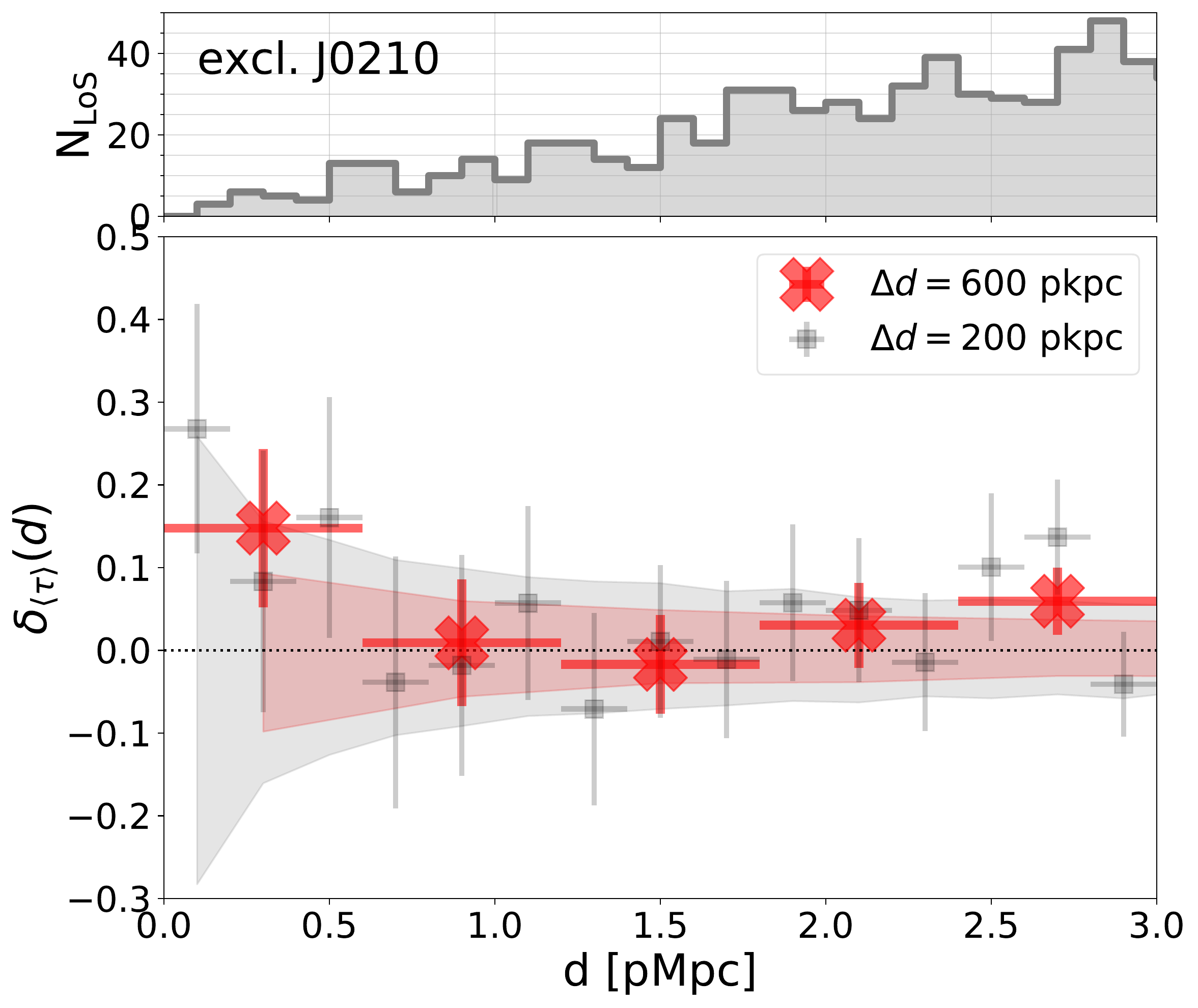}
    \caption{The fluctuation of the average \optdep~as a function of distance to LAEs, $\delta_{\left<\tau\right>}(d)$, for the case excluding $J0210$.
    The symbols are the same with Figure \ref{fig:average_tau}.
    The $J0210$ does not alter the general trend.}
    \label{fig:average_tau_e1f}
\end{figure}

\singlespace




\bibliographystyle{yahapj}
\bibliography{Bib_2020a}



\end{document}